\documentclass{article}

\usepackage{microtype}
\usepackage{graphicx}
\usepackage{subfigure}
\usepackage{booktabs} 
\usepackage{hyperref}
\usepackage[ruled,vlined]{algorithm2e}
\usepackage{multirow}
\usepackage[accepted]{icml2025}
\usepackage{amsmath}
\usepackage{amssymb}
\usepackage{mathtools}
\usepackage{amsthm}
\usepackage[capitalize,noabbrev]{cleveref}

\usepackage{multirow}
\usepackage{booktabs}   
\usepackage{adjustbox}  
\theoremstyle{plain}

\theoremstyle{definition}

\theoremstyle{remark}

\usepackage[textsize=tiny]{todonotes}

\newcommand{\eqnref}[1]{\text{Eq.}\ (\ref{#1})}

\icmltitlerunning{Whitened CLIP as a Likelihood Surrogate of Images and Captions}

\begin{document}

\twocolumn[
\icmltitle{Whitened CLIP as a Likelihood Surrogate of Images and Captions}

\icmlsetsymbol{equal}{*}

\begin{icmlauthorlist}
\icmlauthor{Roy Betser}{technion}
\icmlauthor{Meir Yossef Levi}{technion}
\icmlauthor{Guy Gilboa}{technion}
\end{icmlauthorlist}

\icmlaffiliation{technion}{Viterbi Faculty of Electrical and Computer Engineering, Technion - Israel Institute of Technology, Haifa, Israel.}

\icmlcorrespondingauthor{Roy Betser}{roybe@campus.technion.ac.il}
\icmlcorrespondingauthor{Meir Yossef Levi}{me.levi@campus.technion.ac.il}
\icmlcorrespondingauthor{Guy Gilboa}{guy.gilboa@ee.technion.ac.il}

\icmlkeywords{Machine Learning, ICML}

\vskip 0.3in
]

\printAffiliationsAndNotice{}  

\begin{abstract}
Likelihood approximations for images are not trivial to compute and can be useful in many applications. We examine the use of Contrastive Language-Image Pre-training (CLIP) to assess the likelihood of images and captions. We introduce \textit{Whitened CLIP}, a novel transformation of the CLIP latent space via an invertible linear operation. This transformation ensures that each feature in the embedding space has zero mean, unit standard deviation, and no correlation with all other features, resulting in an identity covariance matrix. We show that the whitened embeddings statistics can be well approximated as a standard normal distribution, thus, the log-likelihood is estimated simply by the square Euclidean norm in the whitened embedding space. The whitening procedure is completely training-free and performed using a pre-computed whitening matrix, hence, is very fast. We present several preliminary experiments demonstrating the properties and applicability of these likelihood scores to images and captions. Our code is available \href{https://github.com/rbetser/W_CLIP/tree/main}{here}.

\end{abstract}

\section{Introduction}
\label{introduction}

Computing likelihoods for images is a challenging yet valuable task with numerous applications in computer vision, such as image generation \citep{ramesh2022hierarchical} and editing \citep{kawar2023imagic}. Traditional approaches, including diffusion models, primarily rely on the likelihood gradient or score function, limiting direct likelihood computation \citep{ho2020denoising, song2020denoising}. 


\begin{figure}[htbp]
    \centering
    \includegraphics[width=0.4\textwidth]{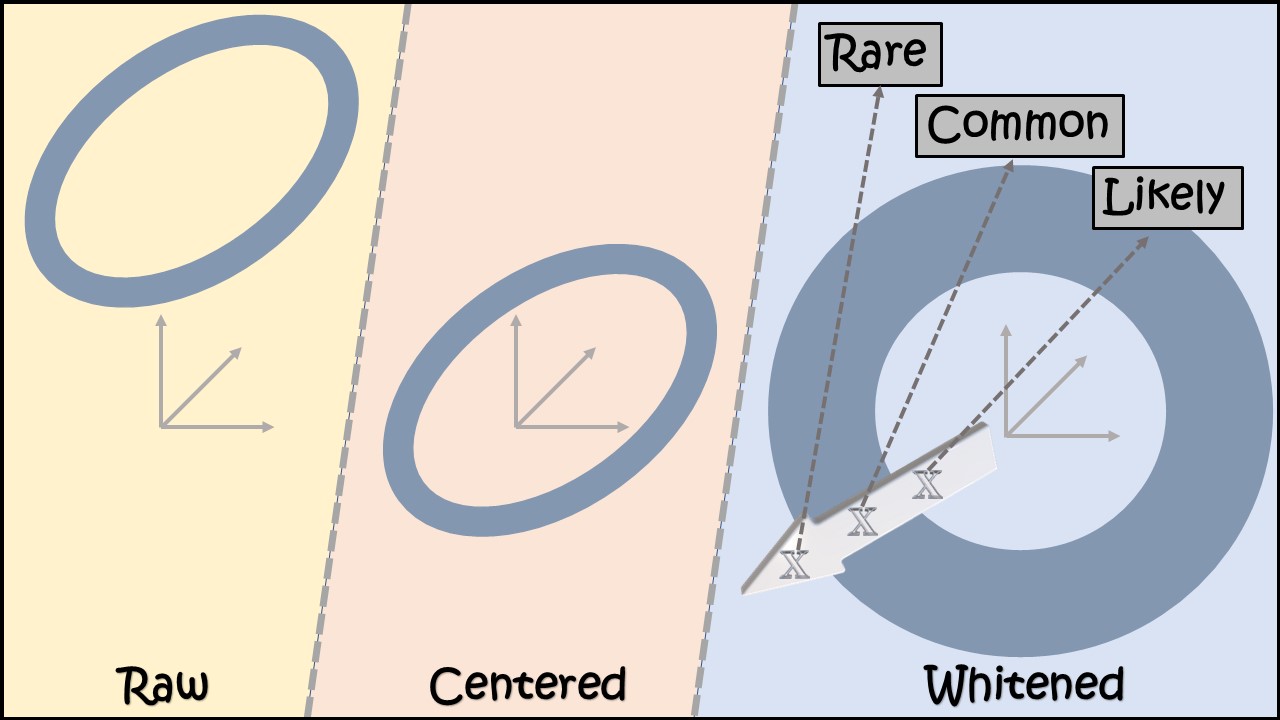}
    \caption{\textbf{Raw, centered and whitened CLIP geometry.} The whitened CLIP space is isotropic, transforming the original ellipsoid shaped space into an hypersphere. In this space, the embedding norm reflects likelihood level. Higher norms correspond to lower probabilities.}
    \label{fig:sphere and ellipse}
\end{figure}

Contrastive Language-Image Pre-training (CLIP) \citep{radford2021learning}  has become a widely adopted embedding for dual text-image semantics. However, its potential as a likelihood surrogate remains unexplored. This paper introduces \emph{Whitened CLIP} (W-CLIP), a linear whitening transformation of the CLIP latent space, where each feature is standardized to have zero mean and identity covariance. In this whitened space, 
we validate by statistical tests that the 
embeddings approximate normal distribution, hence negative log-likelihood estimations are a function of the Euclidean norm in the transformed space. 

To the best of our knowledge, this represents the first direct computation of likelihood functions for images and text prompts under the CLIP-learned distribution.
%
%

\begin{figure*}[ht]
    \centering
    \includegraphics[width=0.85\textwidth]{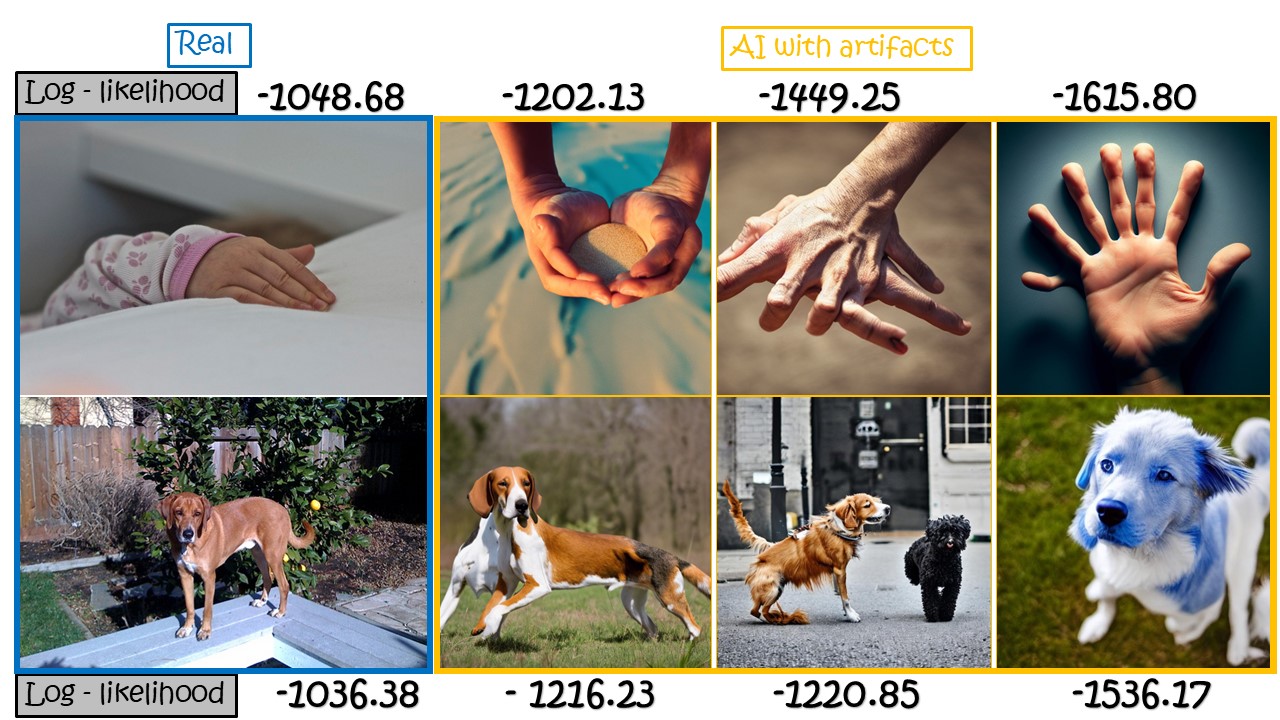}
    \caption{\textbf{Log-likelihood of real and generated images with artifacts.} Real images of a hand and a dog (left) and three similar AI generated images with artifacts. Real images have higher log-likelihood than generated images with artifacts.}
    \label{fig:artifacts}
\end{figure*}

Our main contributions are as follows:
\begin{enumerate}
    \item We propose \textit{Whitened CLIP} (W-CLIP), based on an invertible linear operation, allowing likelihood assessments while retaining the generative and semantic capabilities of CLIP.
    \item We perform quantitative statistical experiments using Anderson-Darling and D’Agostino-Pearson tests, indicating the features in the whitened space can be well approximated by a normal distribution.
    \item We are the first to propose a direct computation of likelihood functions for images and text prompts, under the CLIP learned distribution. For images, to the best of our knowledge, this is the first direct likelihood computation with semantic capabilities.
    \item We show W-CLIP can be used to estimate probability drifts in generative models, discover artifacts in image generation and rank statistical deviation of out-of-distribution (OOD) benchmarks, such as ImageNet-C and ImageNet-R, compared to in-distribution (ID) sets.
    \item For image manipulation, we use W-CLIP to extend Spherical Linear Interpolation (SLERP) by introducing full-circle SLERP, enabling both interpolation and extrapolation between two given images.
\end{enumerate}

\section{Related Work}
\label{related work}

Estimating the likelihood of images, \( P(X) \), is a fundamental task with numerous downstream applications, including super-resolution \citep{li2022srdiff, gao2023implicit}, denoising \citep{tian2020deep, goyal2020image}, and inpainting \citep{yu2018generative, elharrouss2020image}. Early approaches relied on assumptions about natural image smoothness \citep{geman1984stochastic, ruderman1993statistics} and patch distribution 
\citep{zoran2011learning}. Generative models such as Generative Adversarial Networks (GANs) \citep{goodfellow2020generative}, Autoencoders (AEs) \citep{hinton2006reducing, kingma2013auto}, Energy-Based Models (EBMs) \citep{du2019implicit, ou2024energy} and Diffusion models \citep{ho2020denoising, song2020denoising} have further advanced image synthesis by implicitly estimating $P(X)$. However, these methods do not provide explicit access to $P(X)$; for instance, diffusion models approximate the score function, $\nabla_x \log P(X)$, rather than $P(X)$ itself.

In natural language processing (NLP), large language models (LLMs) estimate probabilities directly \citep{devlin2018bert, brown2020language}, while vision-language models (VLMs), including CLIP \citep{radford2021learning} and other recent models \citep{desai2023hyperbolic, chou2024embedding}, embed images and text into a shared space. Despite its success in enabling applications such as captioning \citep{mokady2021clipcap} and image manipulation \citep{kawar2023imagic}, CLIP’s latent space remains underexplored. Known phenomena include the \textit{Narrow Cone Effect}, where embeddings occupy limited angular space \citep{schrodi2024two} and the \textit{Modality Gap}, where image and text distributions are disjoint \citep{liang2022mind, shi2023towards, levi2024double}. \citet{mokady2021clipcap} introduce a mapping network to bridge the modality gap for image captioning.

To the best of our knowledge, this work is the first to analyze CLIP embeddings from a probabilistic perspective and to propose leveraging its latent space as a probability estimator, particularly for the challenging domain of images.


\section{Method: CLIP Likelihoods}
\label{sec: clip likelihood}

\begin{figure*}[ht]
    \centering
    \includegraphics[width=0.85\textwidth]{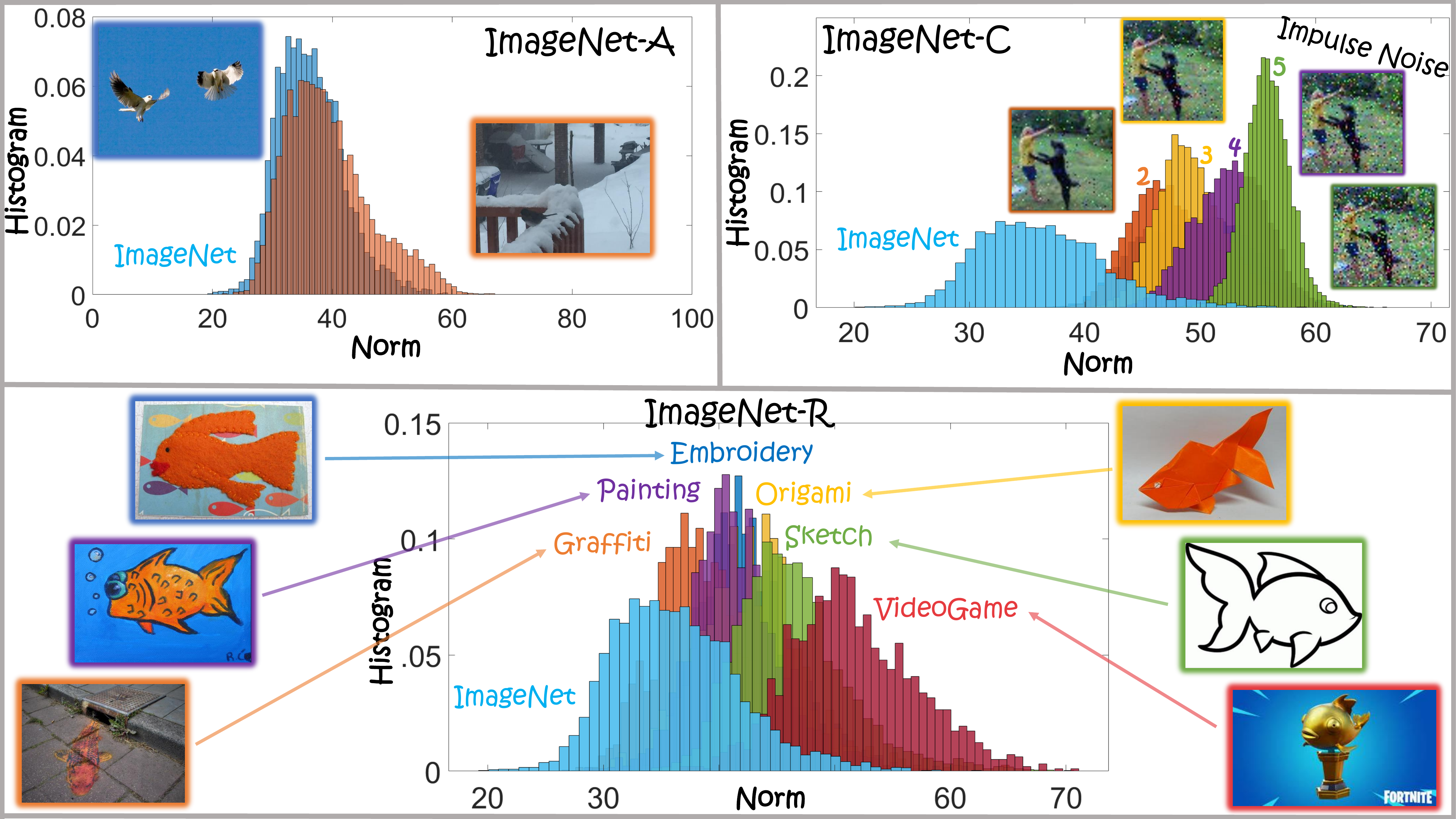
    }
    \caption{\textbf{Norm histograms of ImageNet variations.} Top left: ImageNet-A, comprising of natural adversarial examples, closely aligns with clean ImageNet due to their natural origins. Top right: ImageNet-C histograms under varying impulse noise levels of severity display significantly larger norms than clean ImageNet, indicating distributional deviations. Bottom: ImageNet-R comparison shows that different styles cause varying likelihood shifts, with graffiti closest to real images and video game renditions exhibiting the largest shifts.}

    \label{fig: imagenet}
\end{figure*}

\subsection{Notations}
Let $\mathbf{X} = \{x_1, \cdots x_N\}$ be a set of $N$ random vectors of dimension $d$, $x_i\in \mathbb{R}^d$, where $\mu = \frac{1}{N}\Sigma_1^N{x_i}$ is the mean vector. We denote by $\hat{x}_i
= x_i - \mu$ the centered vector, where
$\mathbf{\hat{X}} = \{\hat{x}_1, \cdots \hat{x}_N\}$. Let $\mathbf{\Sigma} \in \mathbb{R}^{d \times d}$ be the empirical covariance matrix of $X$:
\begin{equation}
    \mathbf{\Sigma} = \frac{1}{N} \mathbf{\hat{X}} \mathbf{\hat{X}}^\top.
    \label{eq:covariance_matrix}
\end{equation}
We recall that the covariance matrix is symmetric, positive semi definite and that the diagonal contains the variance of each feature in the vector.

\subsection{Whitening transform}
\label{sec: whiteneing transfrom}
Given a set of random vectors  with a non-singular covariance matrix \( \mathbf{\Sigma} \), let \( W \in \mathbb{R}^{d \times d} \) be a matrix that satisfies \( W^\top W = \mathbf{\Sigma}^{-1} \). We note that $W$ is not unique. A common way to obtain it is by \emph{principal component analysis} (PCA). Let us diagonalize the covariance matrix $\mathbf{\Sigma} = V\Lambda V^\top$, where \( \Lambda \in \mathbb{R}^{d \times d} \) is a diagonal matrix of the eigenvalues, \( \lambda_i \),  and \( V \in \mathbb{R}^{d \times d} \) consists of the corresponding eigenvectors.  Then, the \textit{whitening matrix} $W$ can be defined as:
\begin{equation}
    W = \Lambda^{-\frac{1}{2}} V^\top.
    \label{eq: w eq}
\end{equation}
Note that $W$ is an invertible matrix. A single vector $x$ is \textit{whitened} by $y = W\hat{x}$ and the \emph{whitened matrix} \( \mathbf{Y} \in \mathbb{R}^{d \times N}\) corresponding to the raw measurement matrix \( \mathbf{X} \) is

\begin{equation}
    \mathbf{Y} = W \mathbf{\hat{X}}.
    \label{eq: whitened_vector}
\end{equation}

$\mathbf{Y}$ consists of \emph{isotropic} random vectors, that is, each vector has zero mean and an identity  covariance matrix ($\mu_Y = 0, \mathbf{\Sigma_Y} = \mathbb{I}$) (see App.~\ref{sec: isotropic}).
The inverse transform from the whitened space to the original space is performed simply by $x = W^{-1}y+\mu$ and in matrix notation $\mathbf{X} = W^{-1} \mathbf{Y}+\mu \cdot \mathbf{1}$, where $\mathbf{1}\in \mathbb{R}^{1 \times N}$ is a row vector of $1$'s.

Given a set of raw CLIP embeddings, The whitening procedure offers three key advantages:
\begin{enumerate}
    \item \( W \) is obtained in a purely data-driven process, without additional meta-parameters.
    \item Since the transform is invertible, all existing applications developed in the raw embedding space can be seamlessly integrated with this approach.
    \item The computation of \( W \) is performed only once and \emph{a-priori}, based on a representative dataset. Memory and computational requirements are very mild, allowing efficient use also in low-resource settings.
\end{enumerate}
It is known that the CLIP latent spaces of images and captions are disjoint \citep{liang2022mind,levi2024double}. Therefore, we treat the distribution of each modality independently.
Additional implementation details and the complete whitening algorithm are in App.~\ref{sec: imp det}, Alg.~\ref{alg:pca_whitening}.

\subsection{Whitened CLIP embeddings}
\label{sec: Whitened CLIP Embeddings}

\begin{figure*}[ht]
    \centering
    \includegraphics[width=0.9\textwidth]{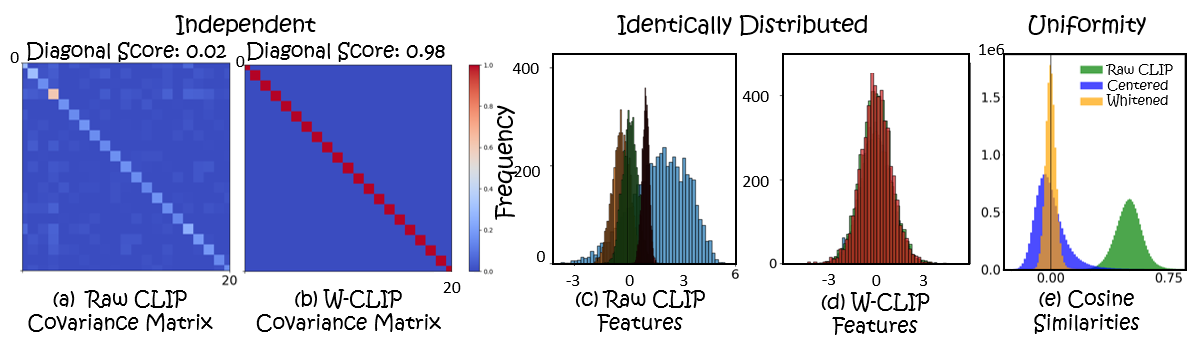}
    \caption{\textbf{Raw CLIP and W-CLIP analytic comparison.} The covariance matrices of raw CLIP (a) and W-CLIP (b) demonstrate the effectiveness of the whitening transformation in achieving unit variance and zero correlation among features. Histograms of four CLIP features (c) vary in mean and variance, whereas four  W-CLIP features (d) exhibit zero mean and unit variance. Cosine similarity histograms for all image pairs (e) across raw, centered, and W-CLIP embeddings reveal that W-CLIP's cosine similarity is concentrated around zero, indicating significantly improved uniformity compared to the centered and raw CLIP spaces.} 

    \label{fig:hists and cov matrices}
\end{figure*}

Our likelihood estimation relies on modeling the whitened CLIP space as   following approximately independent and identically distributed (i.i.d) standard normal distribution. We thus first examine the validity of this approximation. 

\textbf{Normal distribution tests.} To assess how well the whitened embeddings approximate normal distribution, we employ two statistical tests: Anderson-Darling \citep{anderson1954test} and D’Agostino-Pearson \citep{d1973tests}. The Anderson-Darling test evaluates how well the empirical cumulative distribution function (CDF) matches the expected CDF of a normal distribution, placing higher weight on the tails to detect deviations. The D’Agostino-Pearson test combines skewness and shape characteristics measures to assess normality, offering sensitivity to both symmetric and asymmetric deviations. See App.~\ref{sec: normal tests} for additional details regarding these tests, specifically \eqnref{eq: ad defin}, \eqnref{eq: dp defin}.
For stability, the 5000 embeddings of MS-COCO validation set \citep{lin2014microsoft} are divided into 20 equal groups of 250 samples each. As shown in Tab.~\ref{table: normal dist scores}, the results validate that a normal distribution is a good approximation for both image and text embeddings. Specifically, in both tests, more than 90\% of the text features, and more than 98\% of the image features conform to a normal distribution, with average scores that satisfy the test criteria by a large margin. Additional details regarding these tests, empirical statistics, and plots are provided in App.~\ref{sec: normal tests}.  

\textbf{Independent and identically distributed (i.i.d).} Let us break this assumption into independence and identical distributed conditions. For the former, in normal distribution, non-correlation is a sufficient condition for independence. In Fig.~\ref{fig:hists and cov matrices}, the 20 first features of the covariance matrices of raw CLIP embeddings (a) and W-CLIP embeddings (b) are presented. While the CLIP embeddings exhibit correlations between features, the covariance matrix of the whitened embeddings is almost exactly diagonal, indicating that the features are uncorrelated. This is expected, since the whitening transform is designed for exactly this purpose. A metric measuring the proximity of a matrix to being diagonal (in the range $[0,1]$ with $1$ being exactly diagonal) is
\begin{equation}
    \text{Diagonal Score} = \frac{\sum_{i} |\mathbf{\Sigma}_{i,i}|}{\sum_{i,j} |\mathbf{\Sigma}_{i,j}|},
    \label{eq: diag metric}
\end{equation}
where \( \mathbf{\Sigma}_{i,j} \) is an element at row $i$ and column $j$ of the covariance matrix.
 Scores of the full matrices verify that, provided the normal distribution model is valid, the independence assumption holds as well.
Regarding the latter, in Fig.~\ref{fig:hists and cov matrices} the CLIP features exhibit varied mean and variance values (c), while W-CLIP features have all zero mean and unit variance (d), see further results  in Fig.~\ref{fig: mean var}, App.~\ref{sec: normal tests}. Consequently, the whitened embeddings can be approximated reasonably well as i.i.d. features.

\begin{table}[ht]
\centering
\caption{\textbf{Anderson-Darling and D'Agostino-Pearson scores for image and text embeddings.} 
The \textit{Avg.} column  contains the average score for all features, and \textit{Normal Features} represents the percentage of features passing the normal distribution test based on their average score. The threshold score (in brackets) indicates the required condition for normal distribution, with the sign showing whether higher or lower results imply normal distribution.}
\begin{tabular}{|c||cc|}
\hline
 & \textbf{Avg.} & \textbf{Normal Features} \\ \hline \hline
 & \multicolumn{2}{|c|}{\textbf{Anderson-Darling ( $ < 0.752$)}} \\
\textbf{Image}        &   0.4890 &  98.3\%  \\ 
\textbf{Text}         &   0.5926 &  90.1\% \\ \hline \hline
 & \multicolumn{2}{|c|}{\textbf{D’Agostino-Pearson ( $ > 0.05$ )}} \\ 
 \textbf{Image}        &   0.3624 &  99.3\%  \\ 
\textbf{Text}   &   0.2568 &  99.2\%  \\ \hline
\end{tabular}
\label{table: normal dist scores}
\end{table}

\subsection{Log probabilities using W-CLIP}

\begin{figure*}[ht]
    \centering
    \includegraphics[width=0.85\textwidth]{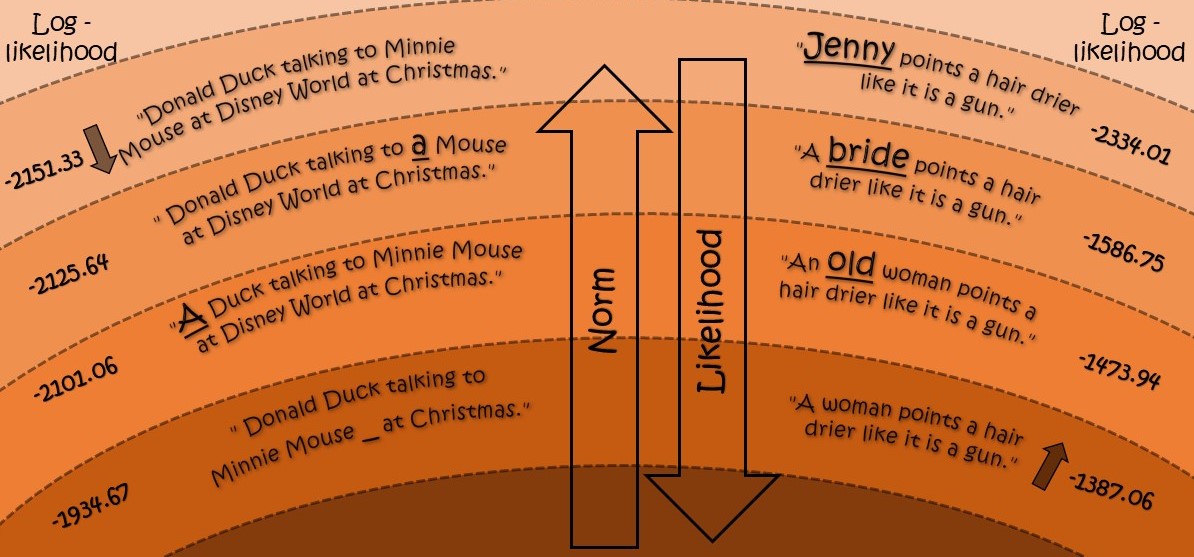}
    \caption{\textbf{Likelihood variation for different levels of details.} 
    The original MS-COCO caption is marked with an arrow, with deviations underlined. Left: Removing details, such as character names or locations, increases likelihood. Right: Adding specificity, such as replacing ``woman'' with ``bride'' or ``Jenny'', decreases likelihood.}

    \label{fig:text prob}
\end{figure*}

\label{sec: log probs}
\textbf{Embedding likelihood.}
The explicit likelihood of a \( d \)-dimensional random vector, \( x \), with i.i.d standard normal variables is:

\vspace{-10pt}
\begin{equation}
    P(x) = \frac{1}{(2\pi)^{\frac{d}{2}}} \exp\left(-\frac{1}{2} \|x\|^2\right),
    \label{eq:isotropic likelihood}
\end{equation}
\vspace{-10pt}

where \( \|x\|^2 = x^\top x \). The log-likelihood is:

\vspace{-10pt}
\begin{equation}
   \ell(x) = \log P(x) = -\frac{1}{2} \left(d\log(2\pi) + \|x\|^2\right).
    \label{eq:isotropic log-likelihood}
\end{equation}
\vspace{-10pt}

Thus, we propose $\ell(x)$ to be an approximation of the log likelihood of image or caption instances, based on the W-CLIP embedding. To the best of our knowledge, this is the first method to directly obtain a probability score for image or text embeddings using CLIP and the first  probability computation for images which is not based on low-level patch statistics but on high-level semantics. In contrast, natural language processing (NLP) language models can directly approximate the negative log-likelihood (NLL) for a text prompt. The relationship between our log-likelihood measure and those of language models is discussed in Sec.~\ref{sec: text model probabilities}.

\textbf{Norm distribution in W-CLIP.}
According to \eqnref{eq:isotropic log-likelihood},
the norm is directly related to the log-likelihood. We thus highlight some consequences and recall the distribution of norms under standard normal statistics.
We first note that the most probable sample resides at the center of the whitened embedding space, nevertheless, the likelihood of sampling this singular point out of the entire space is zero in practice. In general, high-dimensional normal distributions have close to zero mass near the origin. This follows a phenomenon called \textit{Thin Shell} (App.~\ref{sec: thin shell}), which reveals that the majority of the distribution is concentrated near the surface of a sphere of radius $\sqrt{d}$. The chi distribution (\( \chi_d \)), is the appropriate model for the distribution of norms in the whitened space.  We denote the norm of $x$ by \( S = \sqrt{\sum_{i=1}^{d} x_i^2 } \). The log-likelihood of $S$ is:
\begin{equation}
   \log(P(S)) = C(d) + \left( \frac{d}{2} - 1 \right) \log(S^2) - \frac{1}{2} S^2,
\label{eq:log chi}
\end{equation}
where \( C(d) = - \log \left( 2^{(\frac{d}{2}) - 1} \Gamma(\frac{d}{2}) \right) \) and \( \Gamma \) is the Gamma function. The expected value and standard deviation of \( S \) are:

\begin{equation}
    \mathbb{E}[S] = \sqrt{2} \, \frac{\Gamma\left(\frac{d+1}{2}\right)}{\Gamma\left(\frac{d}{2}\right)}, \quad \text{Std}(S) = \sqrt{d - \mu_{S}^2 }.
\label{eq: chi mean std}
\end{equation}



For large \( d \), \( \mathbb{E}[S] \to \sqrt{d - \frac{1}{2}} \). The comparison between theoretical and empirical measurements in Tab.~\ref{table: chi results}, based on MS-COCO, reaffirms the assumed framework of normal distribution. The mean and standard-deviation of the whitened image embeddings closely align with the expected values, while the text embeddings exhibit slightly greater deviation; a trend consistent with the results in Tab.~\ref{table: normal dist scores}.

\begin{figure*}[ht]
    \centering
    \includegraphics[width=0.9\textwidth]{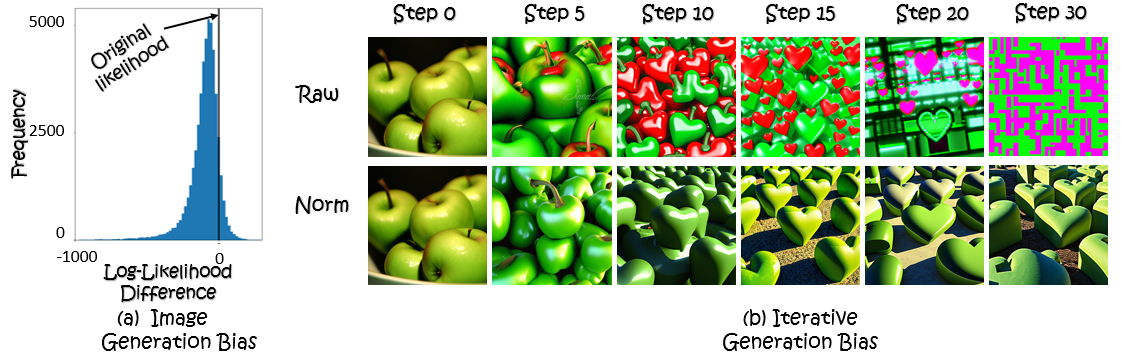}
    \caption{\textbf{Bias in image generation.} 
    Left: Using CLIP-encoded images in MS-COCO validation set as a condition for generating new images. The histogram shows a bias towards lower likelihoods in generated images. Right: Iteratively using UnCLIP to generate images encoded by CLIP with a fixed seed. The raw process gradually becomes noisy (Top), whereas with normalization (to $\sqrt{d}$ at each encoding step), the content drifts but remains within the natural and reasonable image space.} 

    \label{fig: clip loop}
\end{figure*}


\begin{table}[ht]
\centering
\caption{\textbf{Empirical and theoretical measurements.} For $d=768$; relative deviation of the empirical (Emp.) from the theoretical (Theo., \eqnref{eq: chi mean std}) values are shown in brackets.} 
\begin{tabular}{|c||cc|}
\hline  
& \textbf{Mean (Emp. / Theo.)} & \textbf{Std (Emp. / Theo.)} \\ \hline \hline 
\textbf{Image} & $27.43 / 27.7 (0.98\%)$  & $3.94 /3.96 (0.55\%)$ \\
\textbf{Text}  & $28.49 / 27.7 (2.85\%)$  & $5.72 / 6.60 (13.24\%)$\\ \hline
\end{tabular}
\label{table: chi results}
\end{table}
\vspace{-10pt}

\section{Experiments}
\label{sec: experiments}
All the experiments in this section employ the CLIP ViT-L/14 model and utilize the MS-COCO validation set to compute the whitening matrix $W$.

\subsection{Attributes of W-CLIP}
\textbf{Text complexity.} Our observation is that more complex and specific words, such as names, are expected to yield lower likelihood scores. In Fig.~\ref{fig:text prob} we show results of caption editing. Words are replaced with either generic or specific terms, for example, by adding or removing names. The likelihood scores adjust accordingly, decreasing for more specific terms and increasing for more generic ones. Additional examples are available in Figs.~\ref{fig: add det}, \ref{fig: remove det}, App.~\ref{sec: detail text exp}.

\textbf{Uniformity enhancement.} An additional desirable property promoted naturally in the whitened space is \textit{uniformity} \citep{wang2020understanding}.  Fig.~\ref{fig:hists and cov matrices}.e presents a histogram of cosine similarities between all possible image pairs (predominantly composed of negative examples) in the raw, centered and whitened spaces. In the whitened space, cosine similarities are concentrated near zero with smaller variance. In contrast, the centered space exhibits higher variance, while the raw CLIP space has similarities centered around 0.5 with high variance. These results indicate that the whitened CLIP distribution is more uniform.

\subsection{Data Analysis using W-CLIP}
\label{sec: data analysis}
\textbf{Artifact detection.} 
An important attribute of any image likelihood function is its capacity to discriminate between authentic and synthetic images, with particular emphasis on identifying artifacts present in synthetic counterparts. In Fig.~\ref{fig:artifacts}, we compare the likelihood of real images, and AI-generated ones from the SynArtifact dataset \citep{cao2024synartifact} containing notable artifacts. All generated images have lower likelihoods than their real counterparts. Additional examples are provided in Fig.~\ref{fig: more artifacts}, App.~\ref{artifact example}.

\textbf{Domain shift.} Fig.~\ref{fig: imagenet} evaluates a subset of ImageNet \citep{deng2009imagenet}, as presented in \citet{imagenet_loc}, in comparison to ImageNet-A \citep{hendrycks2021natural}, ImageNet-C \citep{hendrycks2019benchmarking}, and ImageNet-R \citep{hendrycks2021many}.
Here we show the distribution of \emph{norms} of each set, instead of the \emph{likelihood} estimation. Following
\eqnref{eq:isotropic log-likelihood}, we have
$$ \|x\| = \sqrt{-2 \ell(x) -d \log(2\pi)},$$
where $\ell(x)$ is the log-likelihood estimation of $x$. Thus it is a simple monotonic transformation, which in some cases may serve as an alternative, more convenient, visualization.
One should notice that a higher norm indicates lower likelihood.
ImageNet-A consists of natural adversarial images. Since the images are natural, their norm distribution is similar to that of ImageNet, apart from a slight shift toward higher values. ImageNet-C introduces common corruptions (e.g., impulse noise), with higher noise levels corresponding to lower likelihoods and distributions consistently below ImageNet. ImageNet-R assesses robustness to domain shifts with renditions like art, graffiti, and video games. Renditions closer to real images, like graffiti, have lower norms than video games, but all renditions exhibit higher norms than ImageNet. For additional Imagenet-C corruptions see Fig.~\ref{fig:imagenet_c}, App.~\ref{sec: more imagenet}.

\subsection{Image manipulations}
\label{sec: image mani}

\begin{figure*}[htbp]
    \centering
    \includegraphics[width=0.89\textwidth]{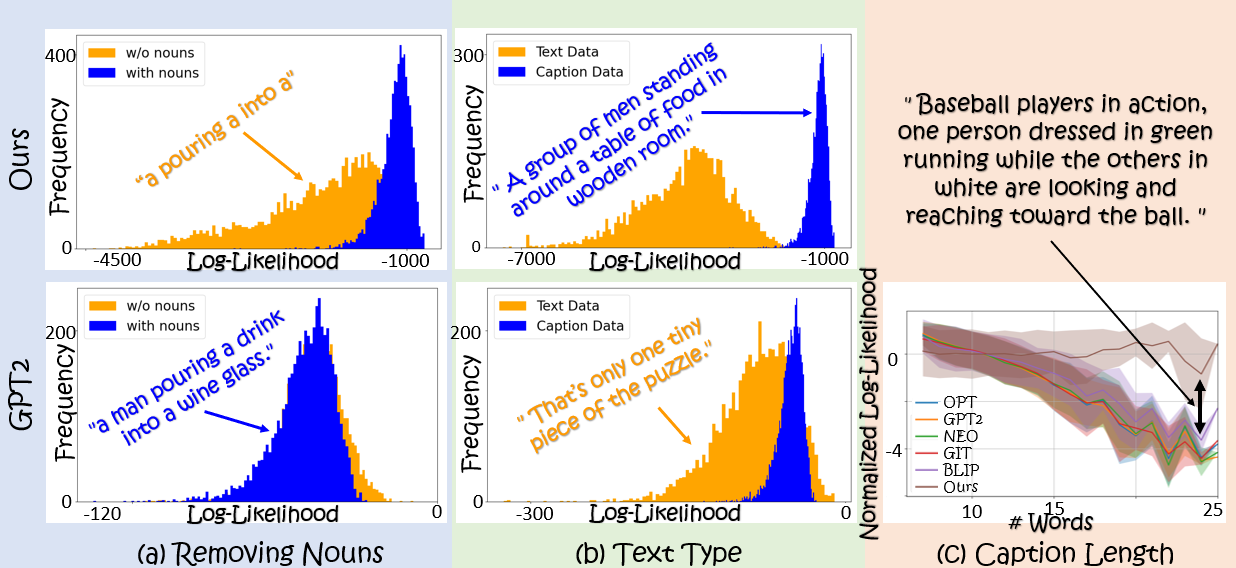}
    \caption{\textbf{Differences between likelihood functions.} 
    Our proposed likelihood estimation is highly sensitive to grammatical errors (a), demonstrated by the removal of all nouns from the captions, and text type (b), where \textit{Text Data} refers to a general text dataset (OpenWebText) and \textit{Caption Data} refers to MS-COCO captions. However, it remains less sensitive to caption length (c). In contrast, language models are highly sensitive to caption length and treat captions as being within the distribution of general text. The removal of nouns from the captions causes only negligible changes to the overall distribution of the model's likelihood. Histograms of all language models are in Figs.~\ref{fig: other text type}, \ref{fig: other text nouns}, App.~\ref{sec: text comp app}.}

    \label{fig: text diff}
\end{figure*}

\textbf{Image generation bias and variance.} 
Generative models may produce outputs which are more likely or less likely than intended. Here we give an example how this
can be quantified, allowing to obtain likelihood-bias and likelihood-variance of a generator, as shown in Fig.~\ref{fig: clip loop}. 
Image generation was performed using UnCLIP \citep{ramesh2022hierarchical}, conditioned by a CLIP embedding.
In this experiment, each image from MS-COCO validation set was encoded using CLIP and subsequently ten images were generated by UnCLIP with different random seeds. A histogram showing the likelihood differences between the original and the generated embeddings is provided in Fig.~\ref{fig: clip loop}.a. The results demonstrate a clear bias towards lower likelihoods. This result indicates that our likelihood approximation method can potentially be leveraged as a generated image detector. Thorough investigation of this task will be conducted in future work.


To further understand this phenomena, we implemented an iterative sequence, starting with a real image encoded by CLIP. The resulting embedding is used to generate a new image, which is re-encoded to CLIP. This iterative process caused the generated images to drift in content, quickly degrading into noise. According to the chi distribution, an embedding is most likely to have a norm of approximately $\sqrt{d}$ (\eqnref{eq: chi mean std}). We use this to normalize each embedding in W-CLIP to a norm of \( \sqrt{d} \) and project back to CLIP space, mitigating this issue. While the iterative process still has a semantic drift, reasonable images were consistently produced. Fig.~\ref{fig: clip loop}.b illustrates this process. Additional experiments are provided in App.~\ref{sec: clip loop exp} (Figs.~\ref{fig: clip loop apple}, \ref{fig: clip loop zebra}).

\textbf{Full circle SLERP.} \citet{ramesh2022hierarchical} propose spherical interpolation (SLERP) on image CLIP embeddings to interpolate between images. SLERP is defined as:

\begin{equation}
    \text{SLERP}(t; \mathbf{E}_1, \mathbf{E}_2) = \frac{\sin((1-t)\theta)}{\sin(\theta)} \mathbf{E}_1 + \frac{\sin(t\theta)}{\sin(\theta)} \mathbf{E}_2,
    \label{eq: slerp equation}
\end{equation}

\begin{figure*}[htbp]
    \centering
    \includegraphics[width=0.8\textwidth]{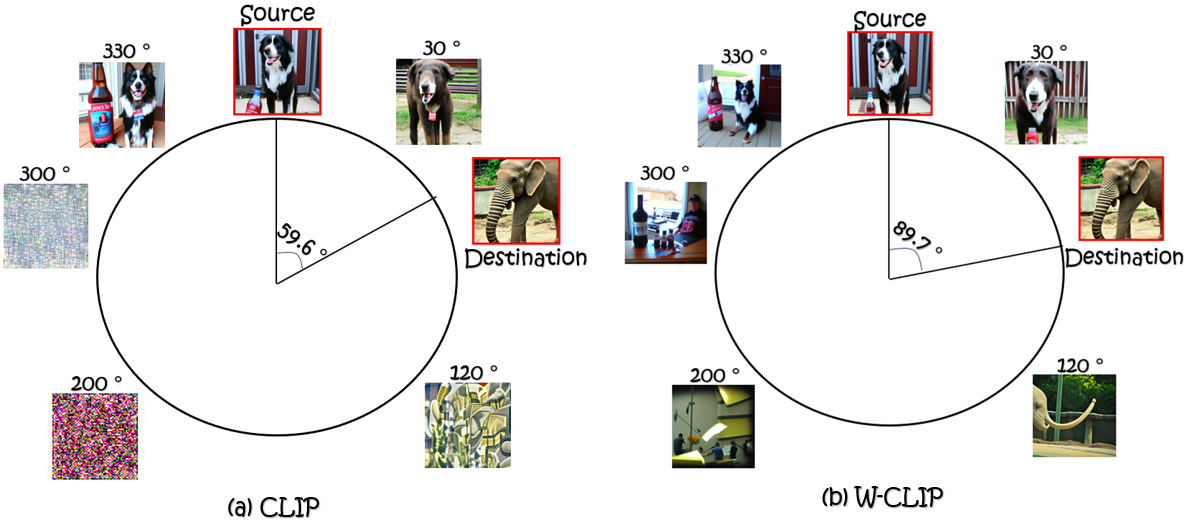}
    \caption{\textbf{Full circle SLERP example.} The full circle SLERP is performed in both the raw CLIP space (a) and in the W-CLIP space (b). The different angle between embeddings in both space is presented. In the raw CLIP space the full circle SLERP results with noise for most of the degrees not between the source and destination embeddings. In the W-CLIP space for all degrees real images are generated.}
    \label{fig: slerp close}
\end{figure*}

where \( \mathbf{E}_1 \) and \( \mathbf{E}_2 \) are embeddings, \( \theta \) is the angle between them (calculated as the normalized dot product), and \( t \in [0, 1] \) is the interpolation step. SLERP assumes embeddings lie on a hypersphere \citep{liang2022mind, wang2020understanding} and is mathematically valid for \( t \) beyond \([0, 1]\). Fig.~\ref{fig: slerp 2d} (App.~\ref{sec: slerp exp}) illustrates SLERP on a 2D circle. When one point is off the circle, interpolation forms an ellipse near the perimeter; if shifted from the origin, the ellipse deviates significantly further. Full-circle SLERP uses an interpolation degree \( \omega \), with \( t = \frac{\omega}{\theta} \) in \eqnref{eq: slerp equation}. In the raw CLIP space, full-circle SLERP often produces noise, with reasonable images only near and between the original embeddings. In the whitened space, it generates consistent images across all angles, with semantic diversity, indicating embeddings remain within the distribution. Images from full circle SLERP examples are in Fig.~\ref{fig: slerp close} and App.~\ref{sec: slerp exp} (Fig.~\ref{fig: slerp far}). In Fig.~\ref{fig: slerp close}, at 300 degrees, which is extrapolation of the source embedding to the reverse direction from the destination embedding, the dog with a bottle of bear (source embedding) becomes a man sitting next to bottles of bear. This is an interesting extrapolation result, not specifically guided. In order to further evaluate this phenomena, and quantify it to quantitative measures we perform an additional experiment. Using MS-COCO validation set, for each image, we performed full-circle SLERP in both the raw CLIP and W-CLIP embedding spaces. In this process, a source image is interpolated toward a destination image along a circular path within the embedding space. Crucially, the image generated at the \(180^{\circ}\) position from the source—referred to as the “opposite image” (generated from the “opposite embedding”)—is invariant to the chosen destination and determined solely by the source. While other positions along the path are influenced by the destination embedding, the opposite embedding is a fixed, symmetric counterpart.

We generate these opposite images using both CLIP and W-CLIP embeddings and observe a stark contrast: in the CLIP space, opposite images degrade into structured noise, whereas in the W-CLIP space, they remain visually natural and semantically meaningful, as shown in Fig.~\ref{fig: slerp close}. The structured noise produced by CLIP exhibits 4×4 pixel blocks and a restricted color palette, suggesting synthetic artifacts. We provide a large visual example in App.~\ref{sec: slerp exp} (Fig.~\ref{fig: slerp opp}).

To quantify these differences, we compute Total Variation (TV), Entropy, and the percentage of extreme saturation values (top or bottom 1\% of the pixel range). All metrics are computed per channel and averaged per image across three sets: original MSCOCO images, CLIP opposites, and W-CLIP opposites. The results are summarized in Tab.~\ref{table: slerp metrics}. These findings confirm that W-CLIP opposites are statistically similar to natural images, whereas CLIP opposites exhibit significantly reduced entropy and variation and much higher percentage of saturation values, indicating a lack of natural structure.

\begin{table*}[ht]
\centering
\caption{\textbf{Cross dataset comparison.} \textit{COCO}: MS-COCO, \textit{F8k}: Flickr8k. \textit{Data T}: dataset used for tests, \textit{Data W}: dataset used to calculate the whitening transform. \textit{Avg. AD, DP}: the average Anderson-Darling, D’Agostino-Pearson p-value test scores (threshold is under 0.752 and above 0.05, respectively). \textit{Correlation} is calculated between likelihood scores on the test data using different whitening matrices.}
\begin{tabular}{|c c||c c c||c c c|}
\hline
& & \multicolumn{3}{c||}{\textbf{Image}} & \multicolumn{3}{|c|}{\textbf{Text}} \\
\textbf{Data T}&\textbf{Data W} & \textbf{Avg. AD} & \textbf{Avg. DP} & \textbf{Correlation} & \textbf{Avg. AD} & \textbf{Avg. DP} & \textbf{Correlation}\\ \hline \hline
\multirow{2}{*}{COCO} & COCO   & 0.489 & 0.362 &  \multirow{2}{*}{0.69} & 0.592 & 0.257 &  \multirow{2}{*}{0.74}\\  
& F8k  & 0.466 & 0.380 &  & 0.574 & 0.282 & \\ \hline   
\multirow{2}{*}{F8k} & COCO  & 0.641 & 0.317  &  \multirow{2}{*}{0.77} & 0.735 & 0.226 & \multirow{2}{*}{0.88} \\  
 & F8k  & 0.522 & 0.329  & & 0.626 & 0.242 & \\  \hline
\end{tabular}
\label{table: flickr comp}
\end{table*}

\begin{table}[ht]
\centering
\caption{\textbf{Opposite image comparison.} Total Variation (TV), Entropy and percentage of saturation pixels (SAT[\%]) for natural images (MS-COCO) and the opposite images generated using CLIP and W-CLIP embedding spaces. W-CLIP represents better the statistics of natural images.}
\begin{tabular}{|l||ccc|}
\hline
\textbf{Method} & \textbf{TV} & \textbf{Entropy} & \textbf{Sat [\%]} \\
\hline
\hline
\textbf{MS-COCO}           & 222.3 & 7.3 & 4.2 \\
\textbf{CLIP}    & 156.7 & 4.8 & 55.5 \\
\textbf{W-CLIP}  & 215.9 & 7.2 & 6.4 \\
\hline
\end{tabular}

\label{table: slerp metrics}
\end{table}
\subsection{Relations to language model probabilities}
\label{sec: text model probabilities}

In natural language processing (NLP), large language models (LLMs) minimize the negative log-likelihood (NLL) during training to learn a probability distribution over sequences. At inference, the NLL is computed by summing the negative log probabilities of each token in the prompt, conditioned on previous tokens. The final NLL is averaged over all tokens, with lower NLL scores indicating higher sequence likelihood under the model’s learned distribution \citep{bishop2006pattern, murphy2012machine}. Our proposed log-likelihood score (\eqnref{eq:isotropic log-likelihood}) approximates the likelihood of text prompts based on a single embedding vector for the entire prompt. We evaluated MS-COCO validation set captions using our method and various language models. Both LLMs (GPT-2 \citep{radford2019language}, NEO \citep{black2021gpt} , OPT \citep{zhang2022opt}) and VLMs (BLIP \citep{li2022blip}, GIT \citep{wang2022git}) were tested. Our method computes log-likelihood values in a different range of values, with correlation values between 0.33 and 0.48 with all language models. See Fig.~\ref{fig: text details}, App.~\ref{sec: text comp app}, for full details. In Fig.~\ref{fig: text diff}, we highlight three main differences between our likelihood score and those of language models:

\begin{enumerate}
    \item \textbf{Caption length.} All language models approximate lower mean likelihood scores as caption length increases. In contrast, our likelihood score is less sensitive to caption length, Fig.~\ref{fig: text diff}.c. \citet{levy2024same} recently demonstrated a degradation in LLM performance on long inputs, particularly in reasoning tasks.
    \item \textbf{Text type.} While text models are trained on general text, CLIP is trained specifically on captions of images. We sampled 5,000 sentences from OpenWebText \cite{Gokaslan2019OpenWeb}, a general text dataset, ensuring that their lengths are comparable to those of MS-COCO captions, and compared both likelihood histograms. For LLMs captions align with the general text distribution, whereas VLMs and our method result in separable distributions (Fig.~\ref{fig: text diff}.b).
    
    \item \textbf{Grammatical errors.} \citet{wu2023chatgpt}  demonstrates that ChatGPT performs poorly on datasets containing grammatical errors, particularly on long sentences. On the other hand, we noticed our method is sensitive to grammatical errors and nonsensical inputs. To test this, we remove all the nouns from the MS-COCO captions and compare likelihood before and after. Our likelihood score is significantly affected, while the language models demonstrate less sensitivity, Fig.~\ref{fig: text diff}.a.
\end{enumerate}

In Figs.~\ref{fig: other text type}, \ref{fig: other text nouns}, App.~\ref{sec: text comp app}, histograms of all other language models are available. In Table~\ref{table: text comp}, we quantify the separation between the likelihoods of different data types and captions with and without nouns. We employ the AUC metric, as defined in ~\eqnref{eq: auc} (App.~\ref{sec: text comp app}), to evaluate the separation between distributions. 
Additional text examples are provided in Fig. \ref{fig: text diff exp}, App.~\ref{sec: text comp app}. It is shown that the likelihood approximated using W-CLIP positively correlates with language model likelihoods but contains unique information derived from CLIP's learned distribution.

\begin{table}[ht]
\centering
\caption{\textbf{Likelihood separation with grammatical errors and different text types.} 
AUC values indicating the separation between likelihood distributions. \textit{Type} compares the separation between captions and general text prompts, while \textit{Nouns} compares the separation between original captions and the same captions with nouns removed. Vision-language models (VLMs) show a high separation for different text types and slightly higher separation when removing nouns compared to language models (LLMs). Our method yields the best separation, especially for \emph{Nouns}.}
\begin{tabular}{|p{0.8cm}||p{0.7cm} p{0.7cm} p{0.7cm} p{0.7cm} p{0.7cm} p{0.7cm}|}
\hline
& \multicolumn{3}{c}{\textbf{LLMs}} & \multicolumn{2}{c}{\textbf{VLMs}} & \multirow{2}{*} {\textbf{Ours}} \\
& \textbf{GPT2} & \textbf{OPT} & \textbf{NEO} & \textbf{BLIP} & \textbf{GIT} & \\ \hline \hline
\textbf{Type} & 0.8   & 0.8 & 0.77 &  0.92 & 0.97 & \textbf{0.999} \\  
\textbf{Nouns} & 0.43   & 0.58 & 0.58 &  0.66 & 0.69 & \textbf{0.94} \\\hline
\end{tabular}
\label{table: text comp}
\end{table}

\subsection{Data generalization}
\label{sec: dataset}
As W-CLIP is completely data-driven we test its generalization capabilities. Flickr8k \citep{hodosh2013framing}, similarly to MS-COCO, is a benchmark for image-captioning tasks that emphasizes real-world imagery and descriptive diversity. In Tab.~\ref{table: flickr comp}, we compare results using MS-COCO and Flickr8k as both the whitening and testing datasets. We evaluate the normal distribution test scores (Anderson-Darling and D'Agostino-Pearson) as in Sec.~\ref{sec: Whitened CLIP Embeddings}, and the correlation of likelihoods computed for the same data, using different datasets for whitening. The results show that whitening with one dataset and testing on another yields similar normal distribution test scores for features, and moderate to high correlations between likelihoods. Additional ablation studies, including different dataset size and utilizing a different CLIP model, are provided in App.~\ref{sec: ablation}. These findings confirm that, although W-CLIP is data-driven, it generalizes well across datasets within the same domain. However, as shown in Fig.~\ref{fig: imagenet}, W-CLIP is sensitive to domain shifts.

\section{Conclusion}
\label{conclusion}
This paper introduces Whitened CLIP, transforming the raw CLIP latent space into an isotropic space. Whitened CLIP is statistically verified to approximate well normal distribution with independent and identically distributed (IID) components, and exhibits enhanced uniformity. The key contribution of this work is the proposal of a direct computation of likelihood functions for images and text prompts within the CLIP-learned distribution. Embeddings in the whitened space approximately follow the standard normal distribution, enabling the use of the squared Euclidean norm to estimate log-likelihood. These likelihood functions effectively identify artifacts, domain shifts, and demonstrate sensitivity to the complexity of details in text captions. Biases in generative models can be detected by comparing the likelihood of generated images to those of real images. Furthermore, the introduction of full-circle SLERP in the whitened space facilitates both interpolation and extrapolation between images. We believe the results of this research can further benefit numerous applications. 

\subsection*{Acknowledgements}
We would like to acknowledge  
support by the Israel Science Foundation (Grant 1472/23) and by the Ministry of Science and Technology (Grant No. 5074/22).


\section*{Impact Statement}
This work advances Machine Learning by enhancing the understanding of a foundation model (CLIP) and utilizing it to derive direct likelihood functions for images and captions. Potential societal consequences of our work are related to downstream tasks that may rely on our findings. None of these consequences need to be specifically highlighted here.

\bibliography{references.bib}
\bibliographystyle{icml2025}

\newpage
\appendix
\onecolumn

\section{Reproducibility}
Our code along detailed instructions is available \href{https://github.com/rbetser/W_CLIP/tree/main}{HERE}. The repository includes: 1) Implementation of our method, reproducing most of our experiments, including simple demo notebooks; 2) Whitening matrices; 3) Additional examples beyond those in the paper and in the appendix, specifically video demonstrations of the full circle SLERP.

\section{Artifacts and Domain Shifts Examples}
\subsection{Image artifacts}
\label{artifact example}
\begin{figure}[ht]
    \centering
    \includegraphics[width=0.97\textwidth]{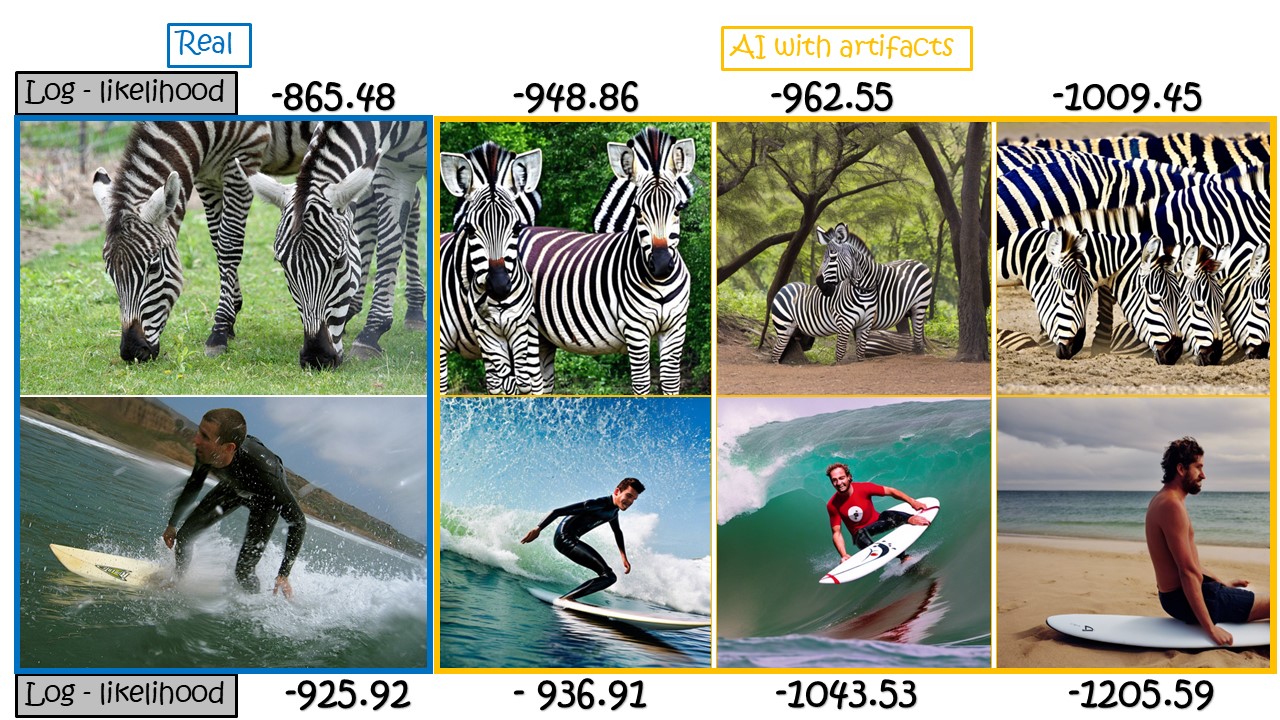}
    \caption{\textbf{Log-likelihood of real and generated images with artifacts.} Real images of zebras and a surfer (left) and three similar AI generated images with artifacts. Real images have higher log-likelihoods than AI generated images with artifacts.}
    \label{fig: more artifacts}
\end{figure}

In Fig.~\ref{fig: more artifacts} we offer additional examples of real images compared to similar generated images with artifacts, as presented in Fig.~\ref{fig:artifacts} in Sec.~\ref{sec: data analysis}.

\subsection{Text artifacts}
\begin{figure}[ht]
    \centering
    \includegraphics[width=0.97\textwidth]{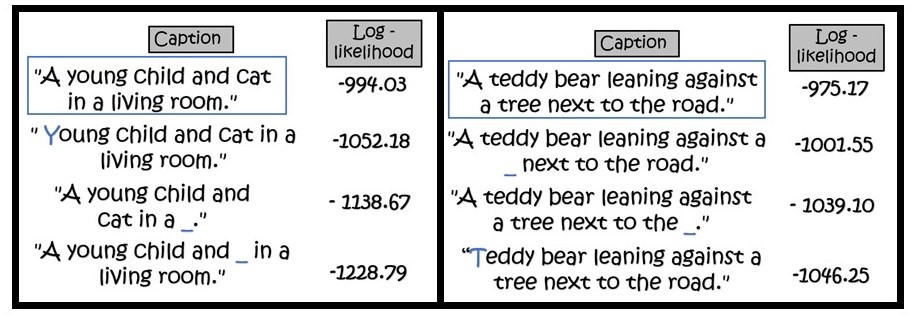}
    \caption{\textbf{Log-likelihood of real captions and captions with artifacts.} Real captions, framed with a blue frame and atrifacted captions, where we removed the first or last words from a caption, or one of the middle words. In all cases the original caption has the highest log-likelihood score.}
    \label{fig: text artifacts}
\end{figure}

Trying to generate artifacts in text captions we remove the first or last words from a caption, or one of the middle words. Examples in Fig.~\ref{fig: text artifacts}. In all cases the original caption has the highest log-likelihood score. 

\subsection{ImageNet datasets}
\label{sec: more imagenet}
\begin{figure}[hp]
    \centering
    \includegraphics[width=0.4\textwidth]{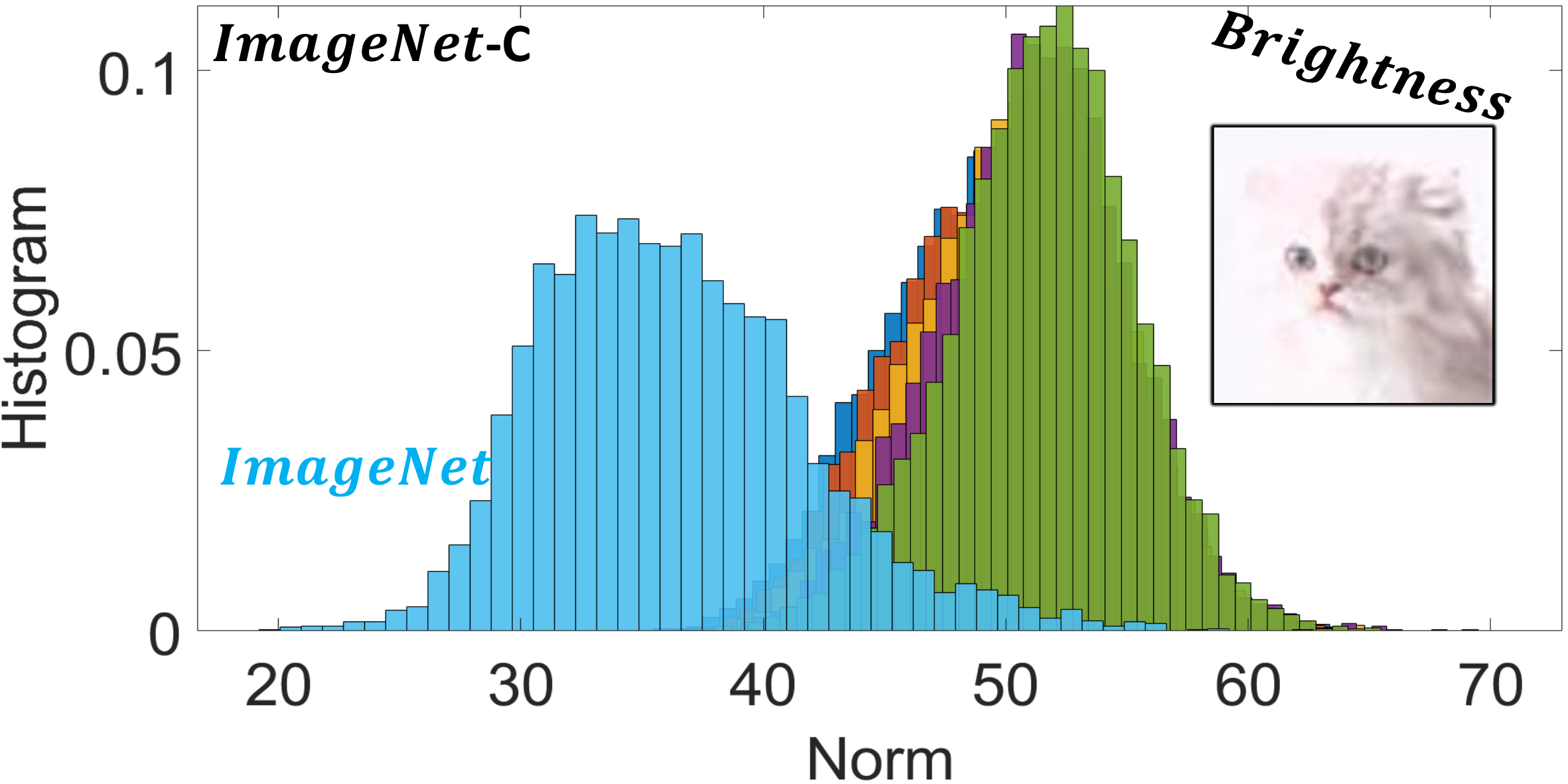}
    \includegraphics[width=0.4\textwidth]{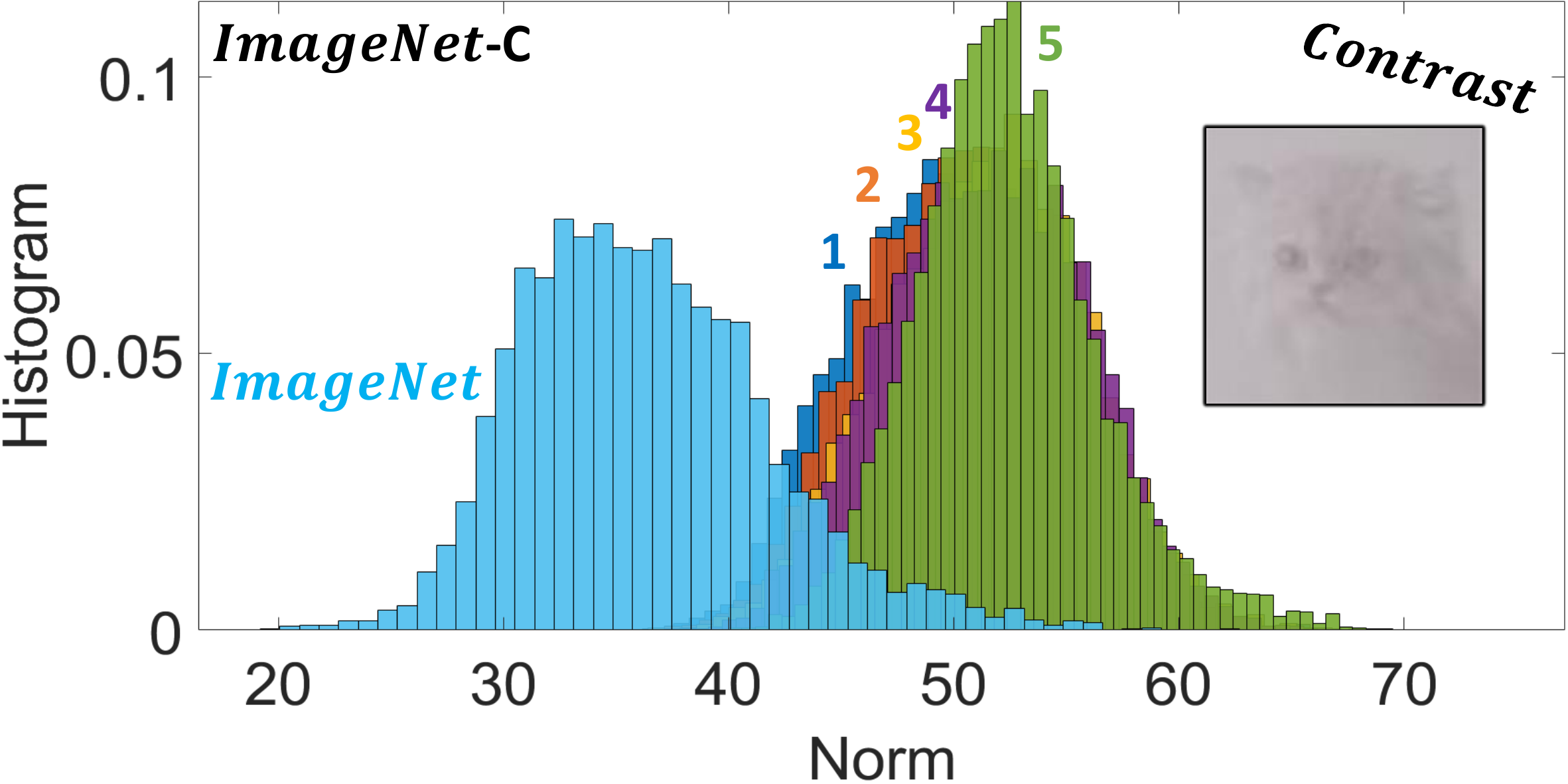}
    \includegraphics[width=0.4\textwidth]{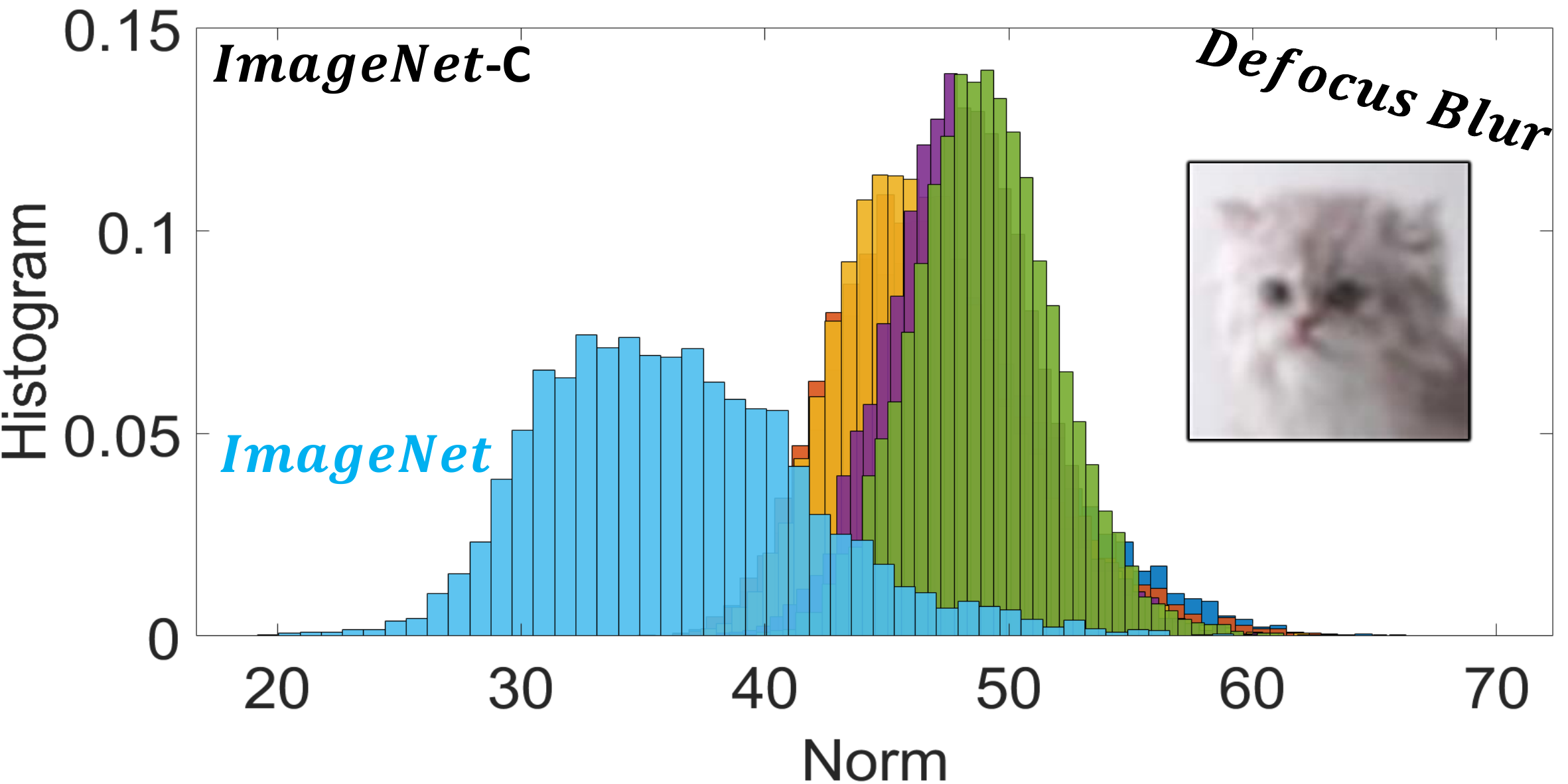}
    \includegraphics[width=0.4\textwidth]{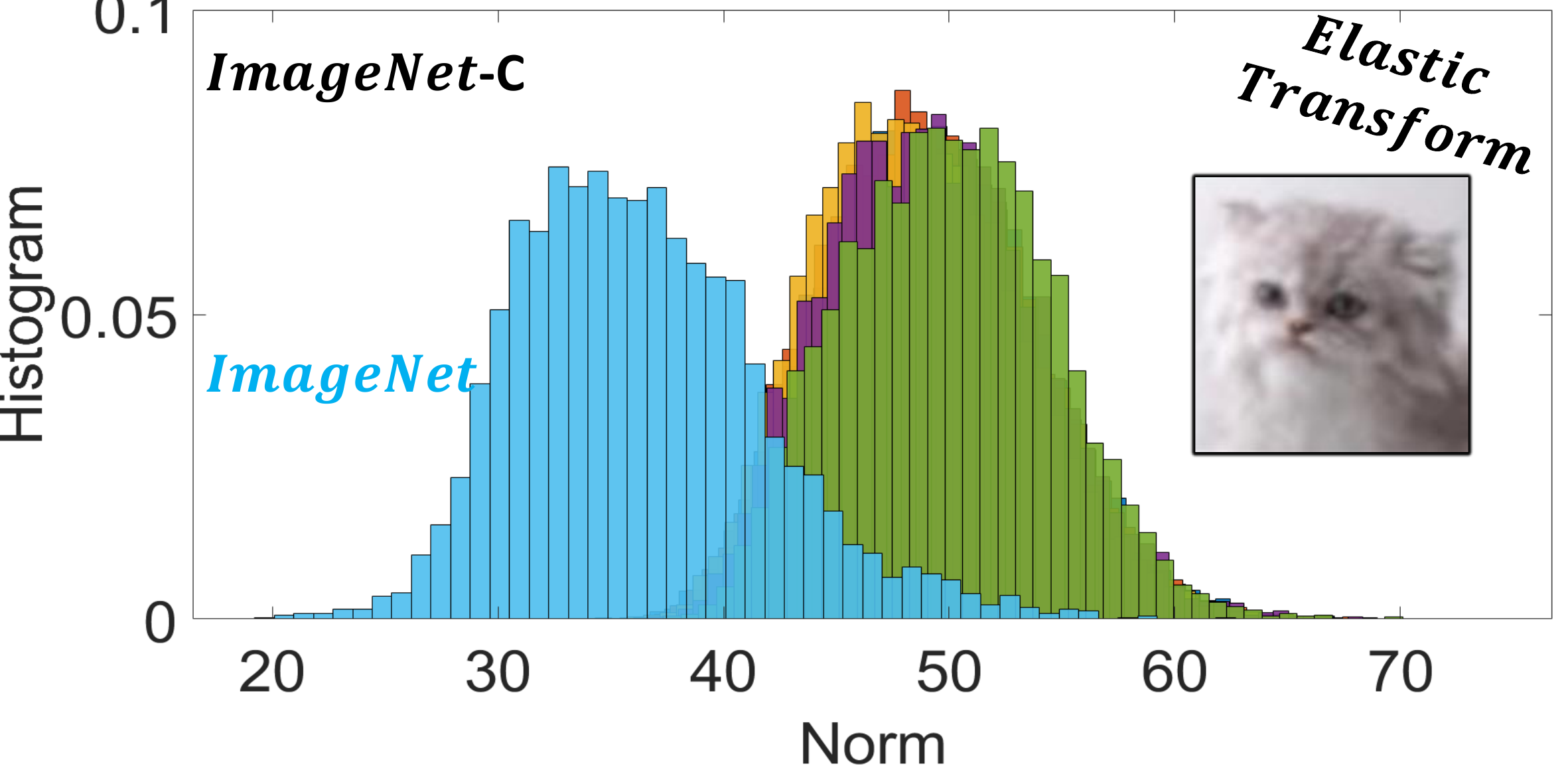}
    \includegraphics[width=0.4\textwidth]{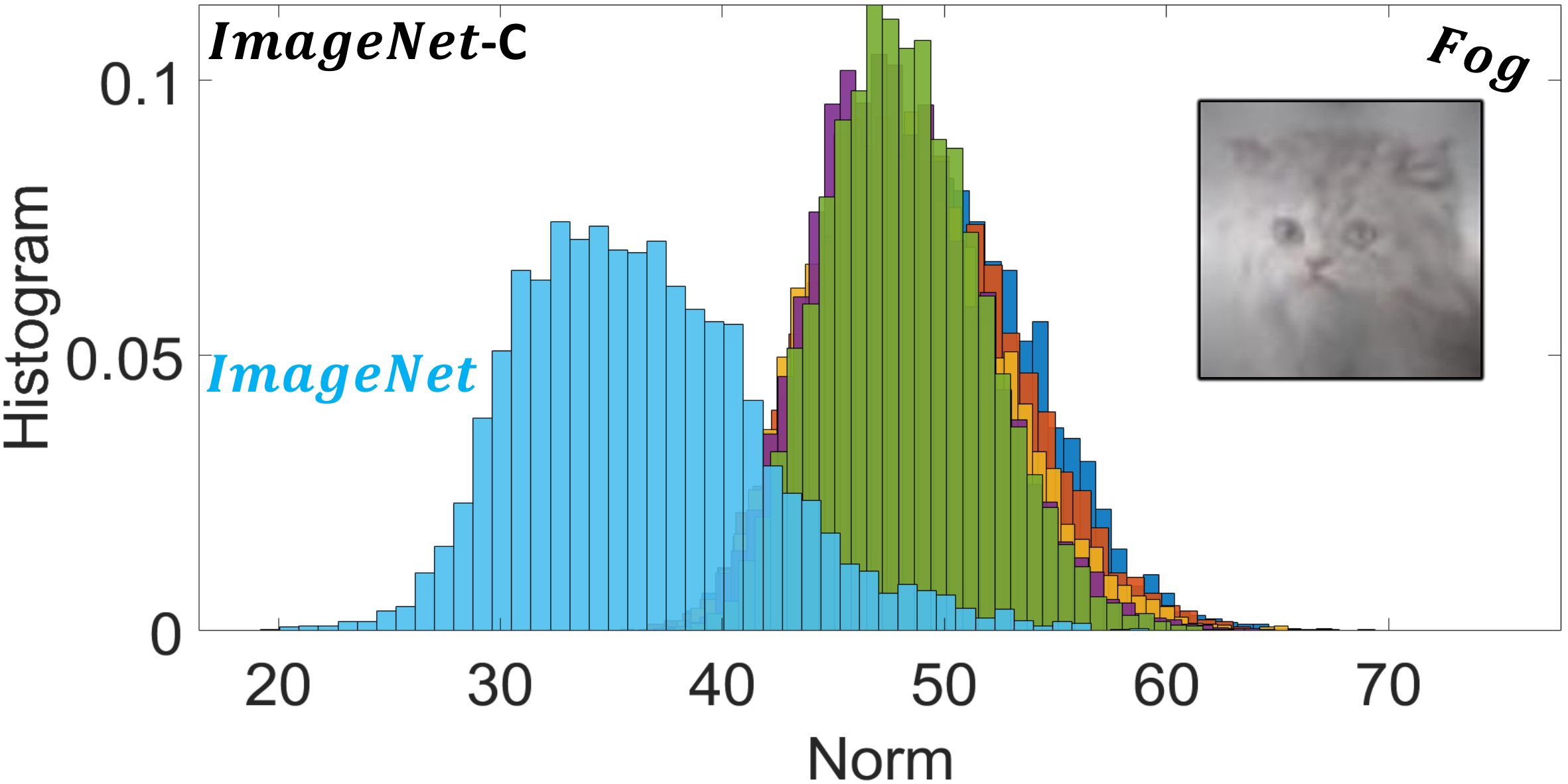}
    \includegraphics[width=0.4\textwidth]{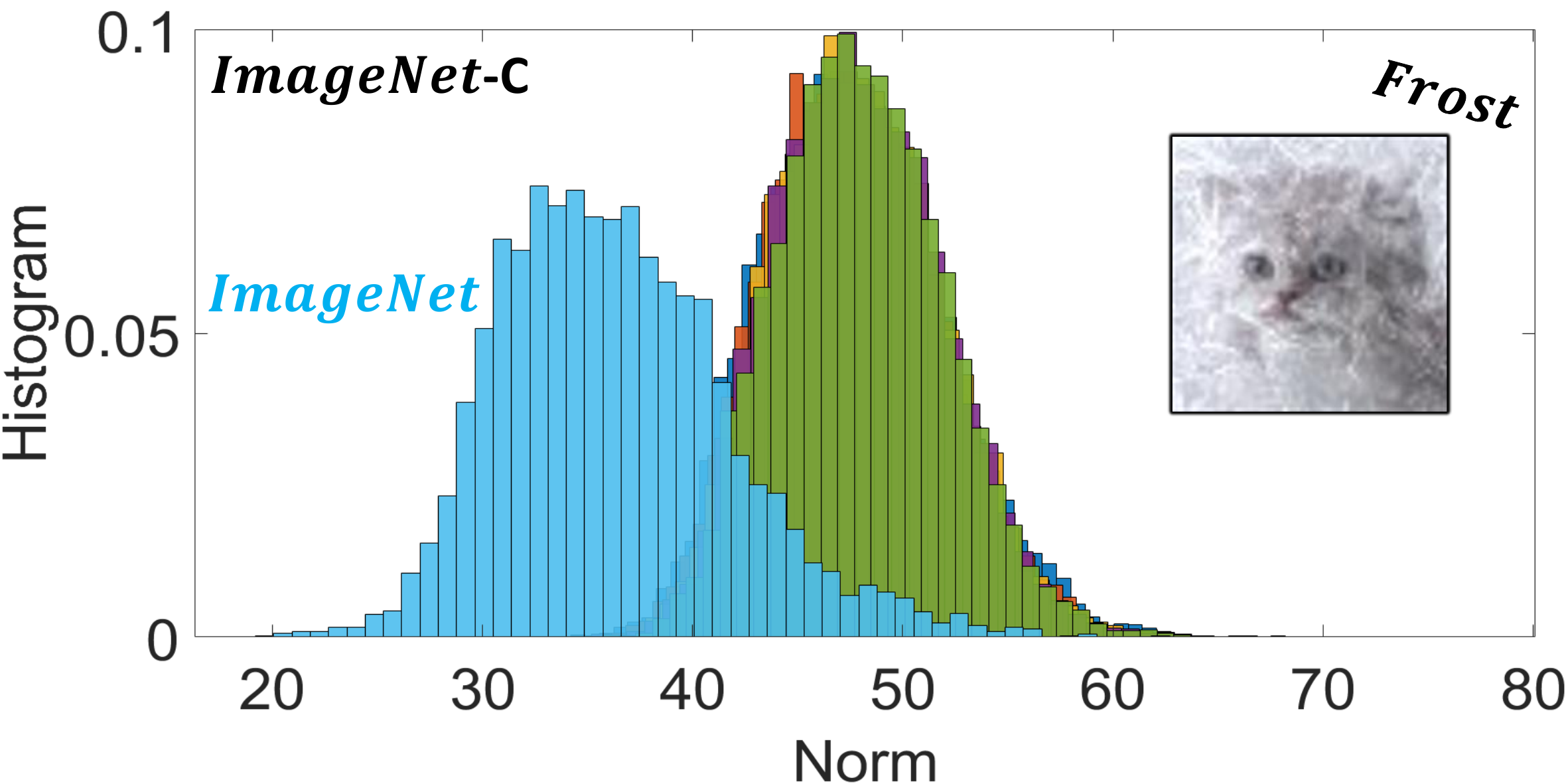}
    \includegraphics[width=0.4\textwidth]{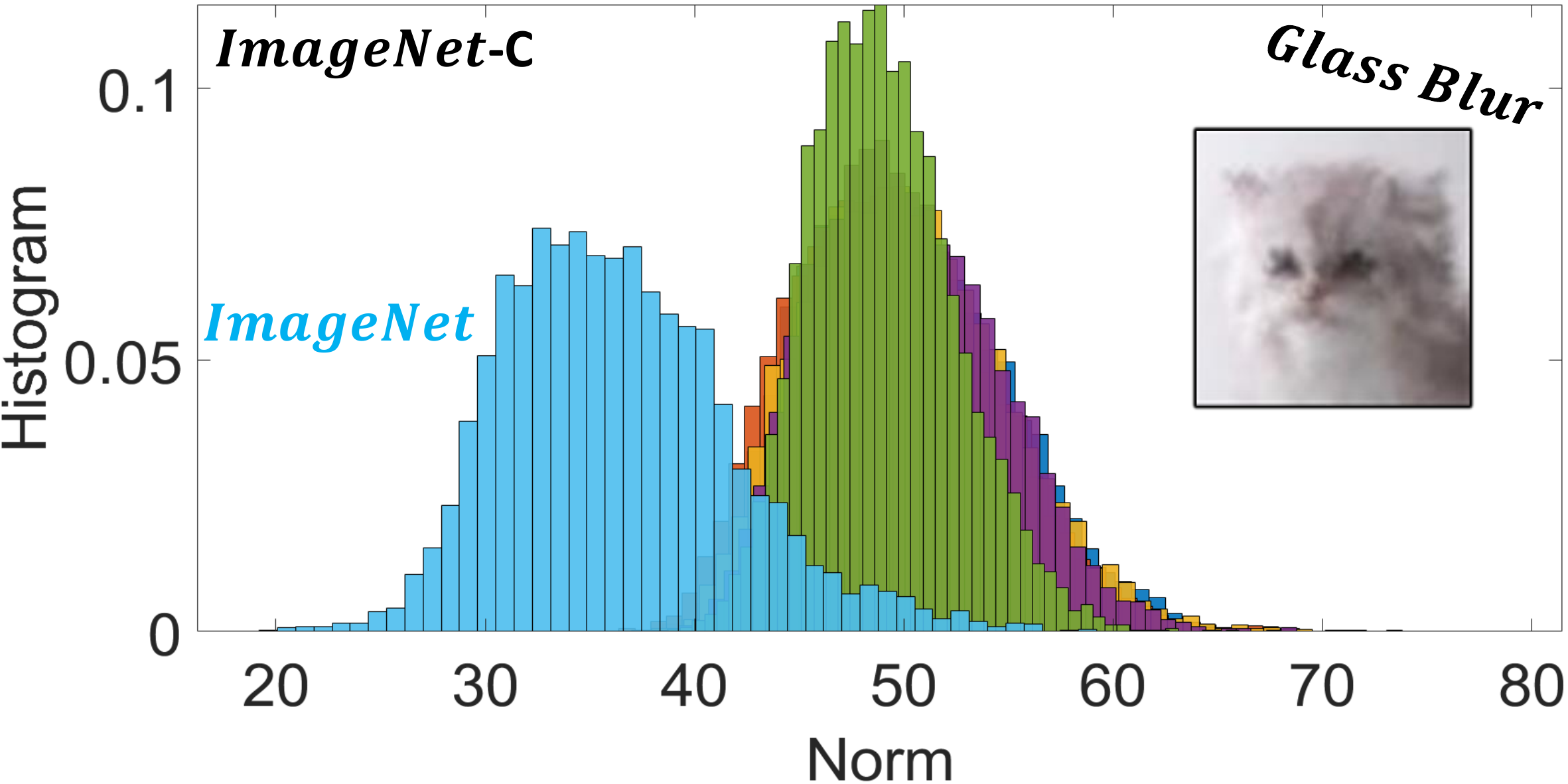}
    \includegraphics[width=0.4\textwidth]{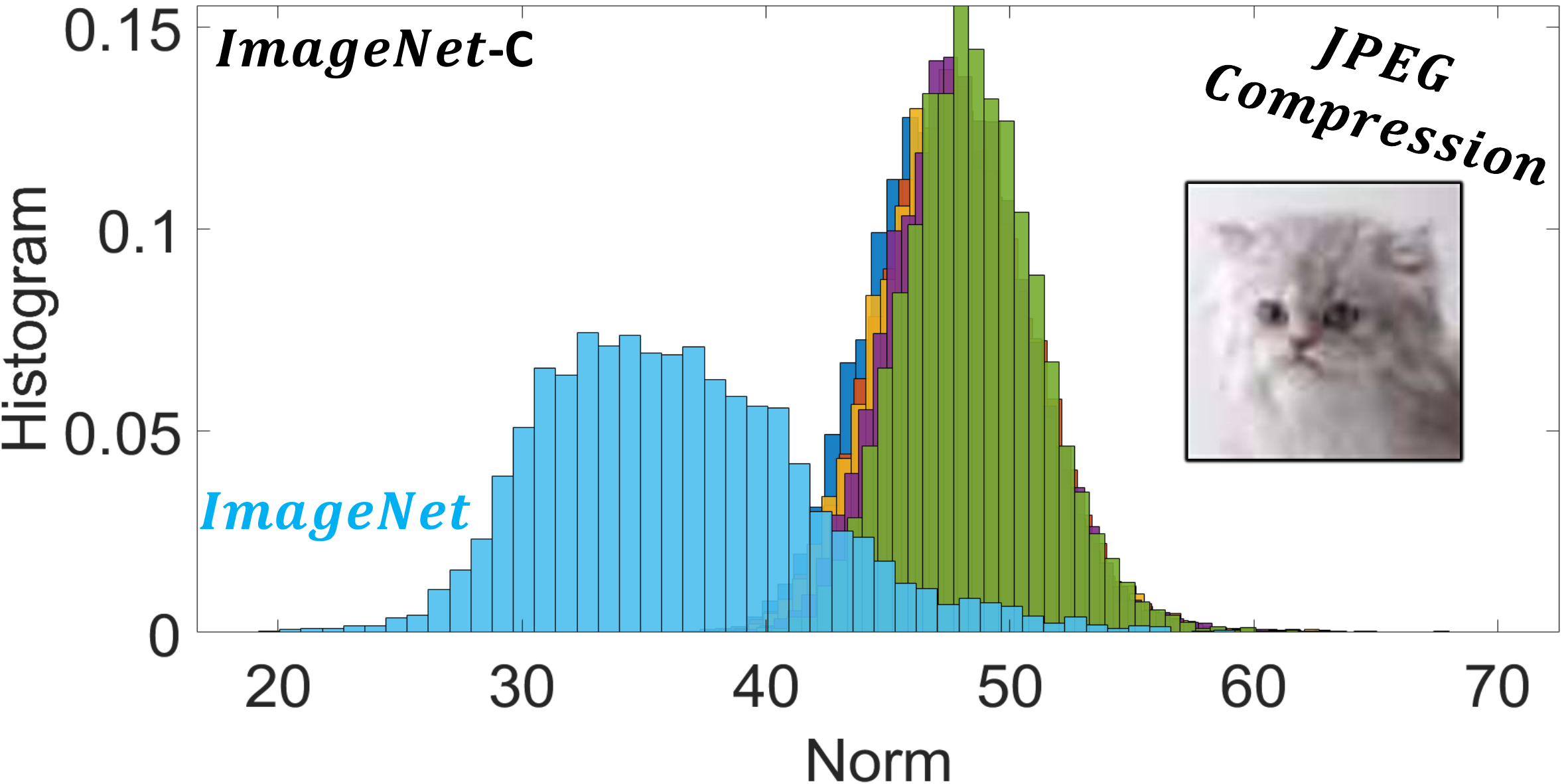}
    \includegraphics[width=0.4\textwidth]{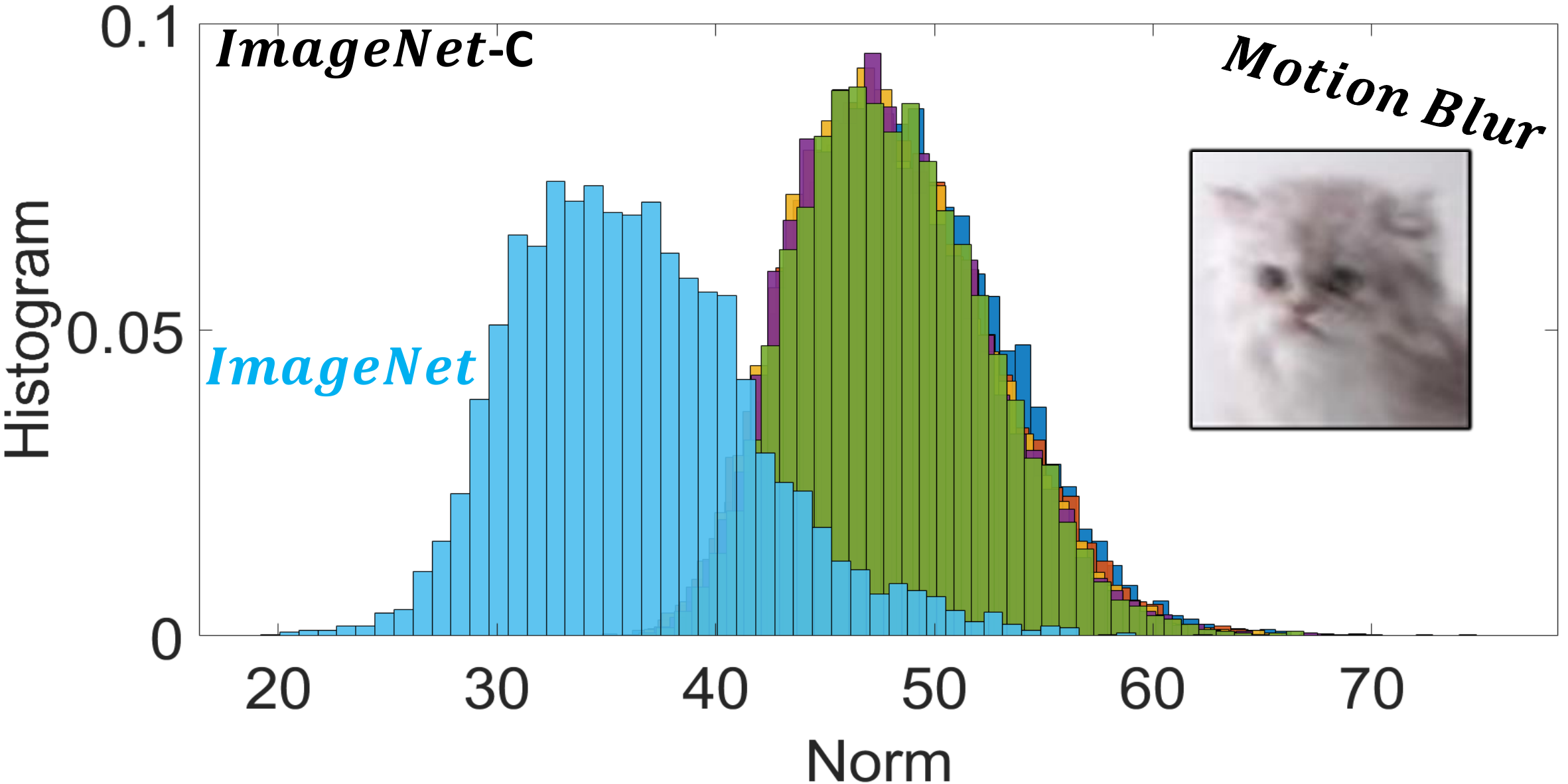}
    \includegraphics[width=0.4\textwidth]{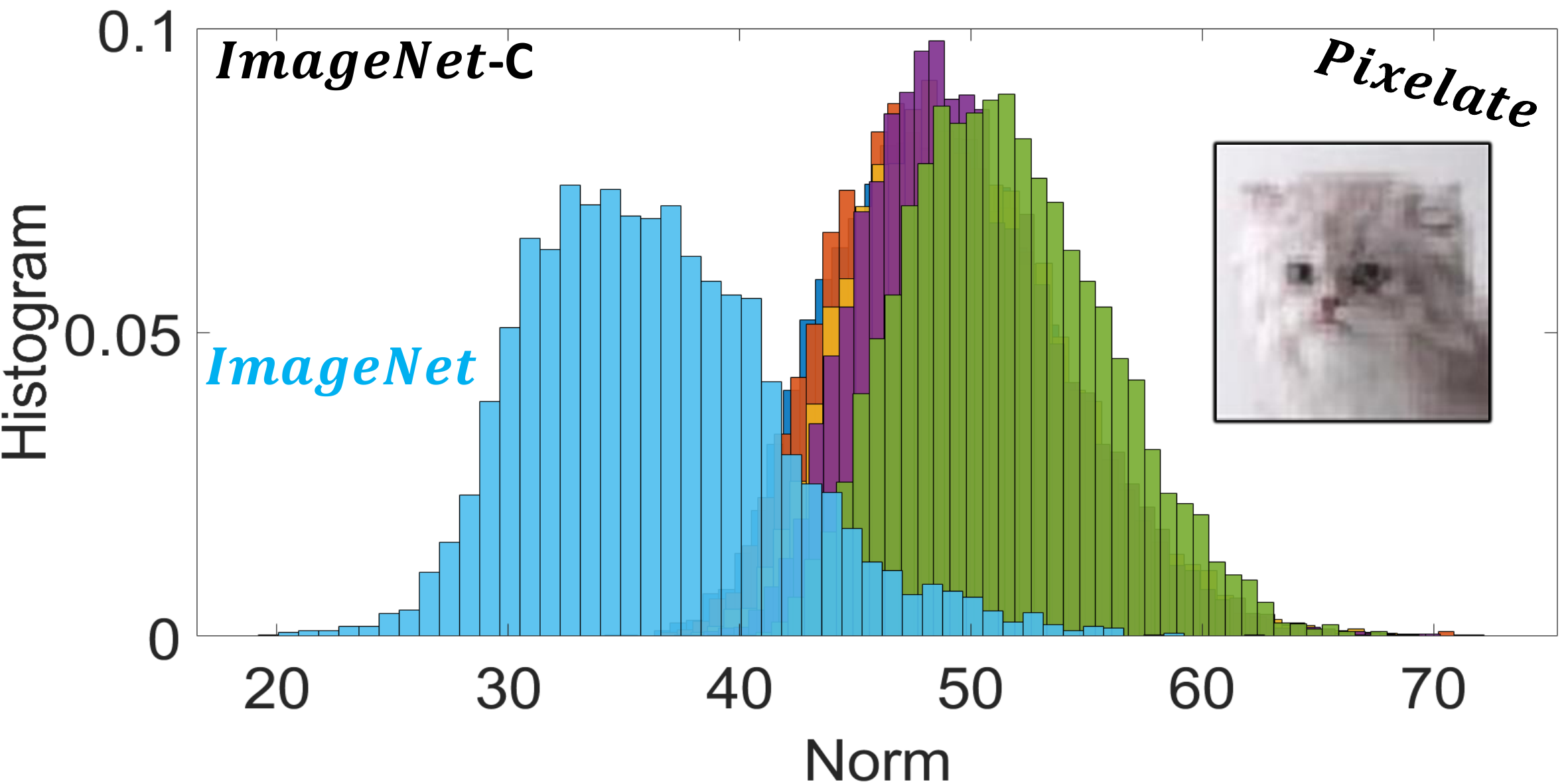}
    \includegraphics[width=0.4\textwidth]{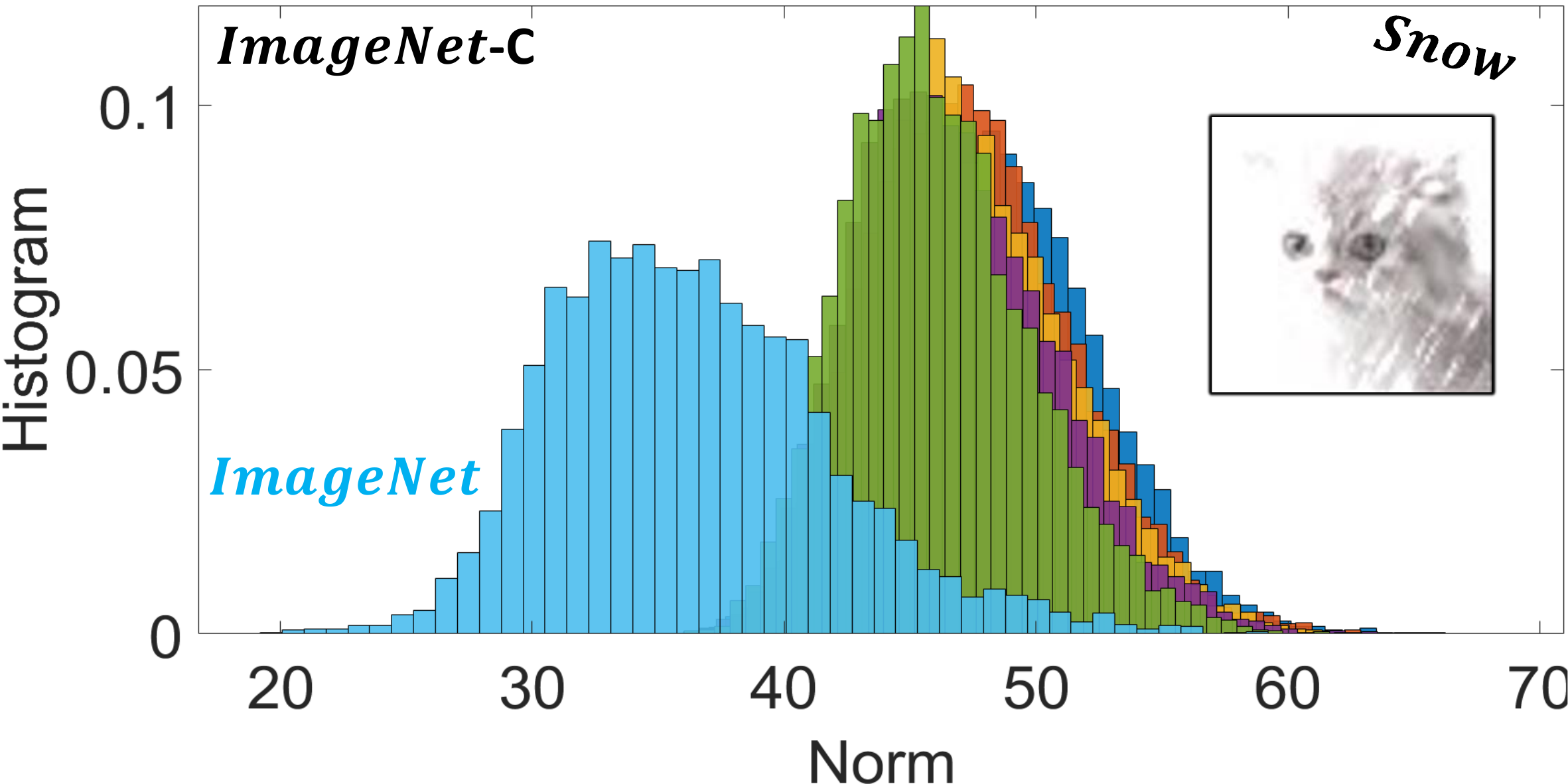}
    \includegraphics[width=0.4\textwidth]{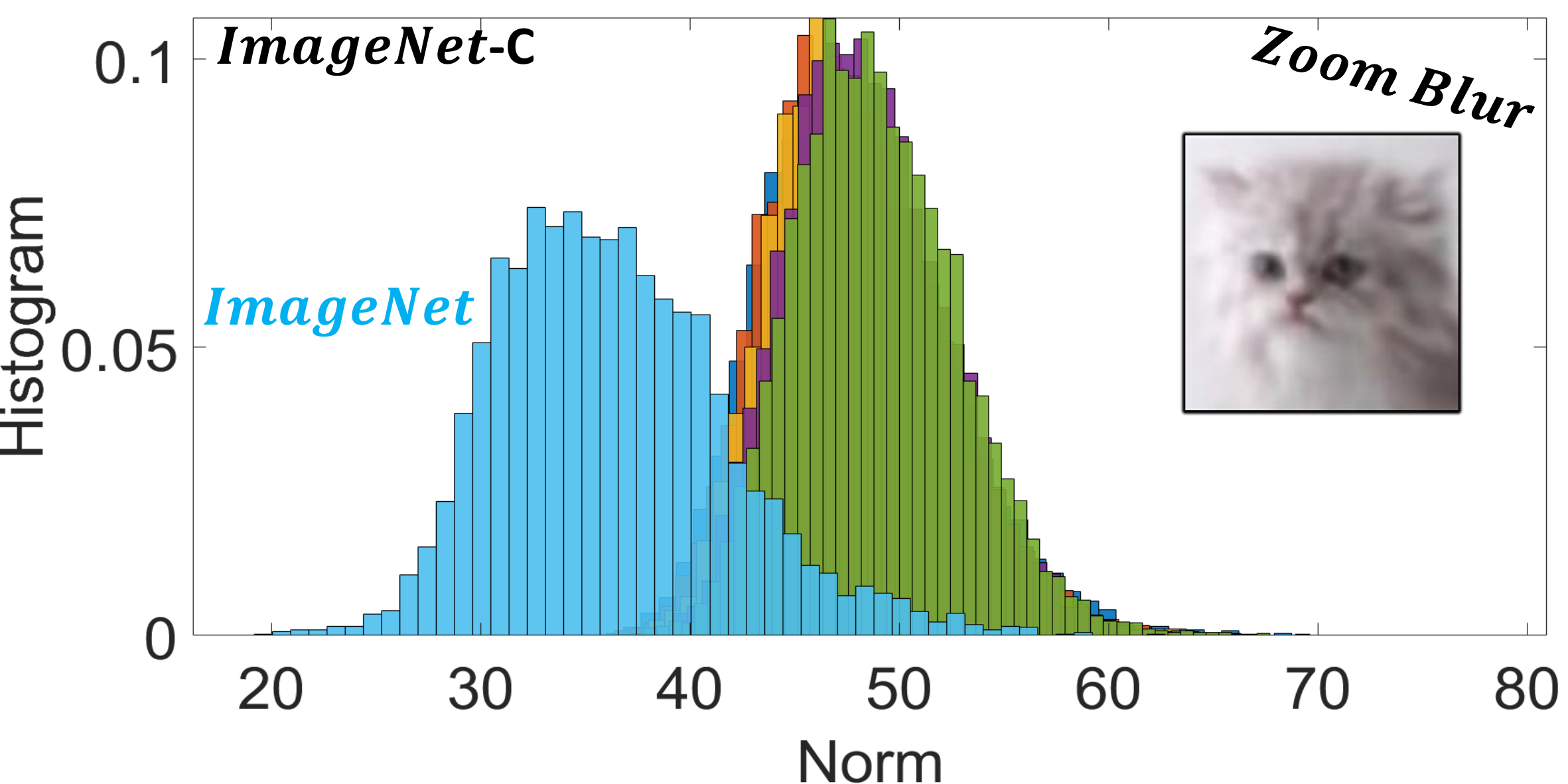}
    \caption{\textbf{ImageNet-C histograms on all corruptions.} All corruptions have a significantly higher norm (lower log-likelihood) than ImageNet. For most corruptions, as level of corruption increases the norm increases. Some corruptions do not show this monotonic behavior (motion/glass blur for example) for different levels of corruption.}
    \label{fig:imagenet_c}
\end{figure}

In Fig.~\ref{fig:imagenet_c} we provide histograms using different corruptions from ImageNet-C. All corruptions have a lower log-likelihood compared to ImageNet.

\newpage

\newpage

\section{Comparison with Language Models Examples}
\label{sec: text comp app}

\begin{figure}[ht]
    \centering
    \includegraphics[width=0.6\textwidth]{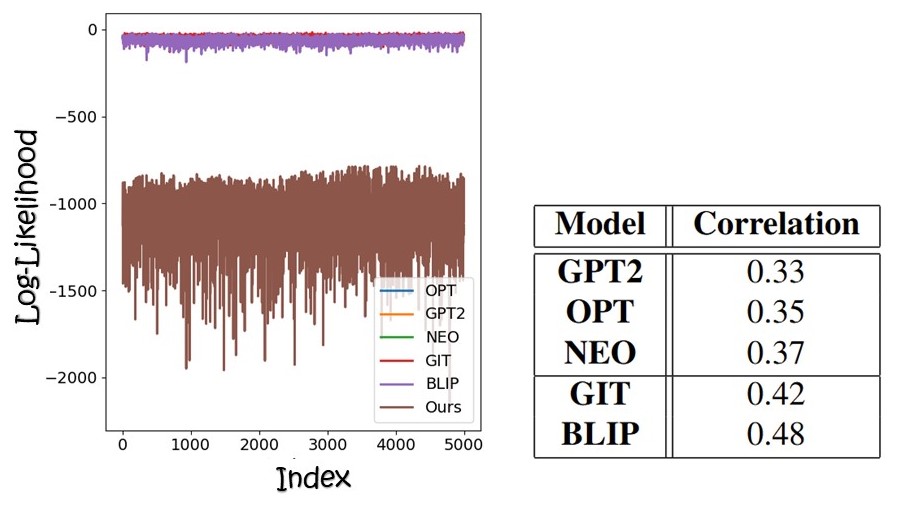}
    \caption{\textbf{Log-likelihood values and correlations.} Log-likelihoods are computed on 5,000 captions from the MS-COCO validation set, with correlations measuring the alignment of each model's log-likelihood and ours.}
    \label{fig: text details}
\end{figure}

In Fig.~\ref{fig: text details}, we present the log-likelihood values for the MS-COCO validation set using language models, along with the correlation between these log-likelihoods and those computed using our method. Our approach yields log-likelihood scores with larger absolute values and greater variance, aligning with its intended design.

\begin{figure}[H]
    \centering
    \includegraphics[width=0.97\textwidth]{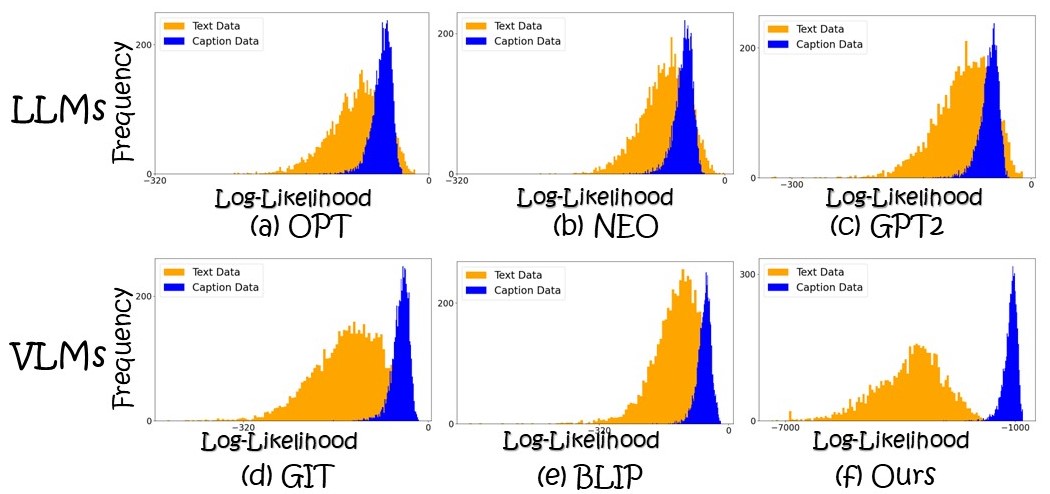}
    \caption{\textbf{Likelihood for different text types.} Comparing likelihood values computed for MS-COCO captions and OpenWebText general text sentences. The sentences from OpenWebText are filtered to have similar lengths to MS-COCO captions. LLMs (OPT, NEO, GPT2) treat captions similarly to general text while VLMs (BLIP, GIT) show some separation. No model shows strong separation like our likelihood does (Tab.~\ref{table: text comp}). }
    \label{fig: other text type}
\end{figure}

\begin{figure}[H]
    \centering
    \includegraphics[width=0.97\textwidth]{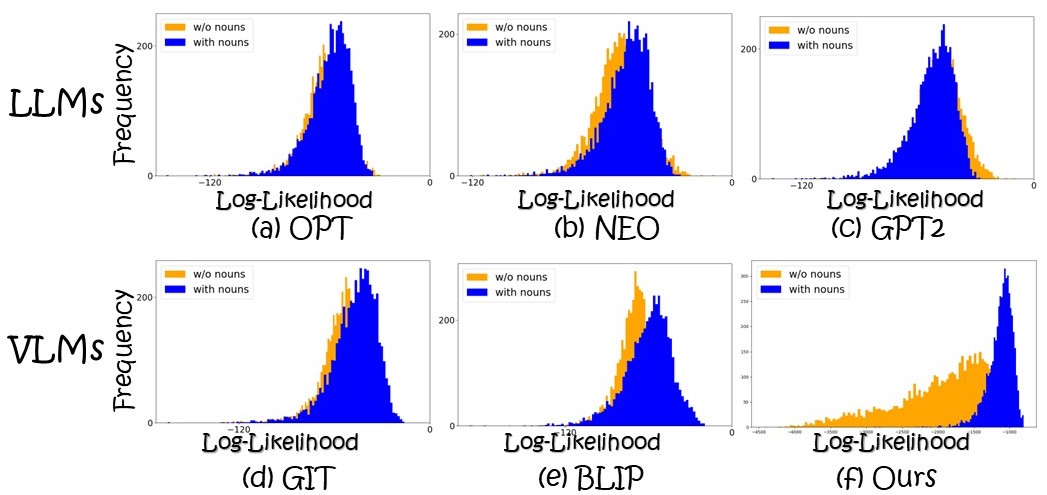}
    \caption{\textbf{Likelihood drift when removing nouns} Comparing likelihood values computed for MS-COCO captions with and without nouns. None of the models show a drift like our likelihood (Tab.~\ref{table: text comp}).}
    \label{fig: other text nouns}
\end{figure}

\begin{figure}[H]
    \centering
    \includegraphics[width=0.97\textwidth]{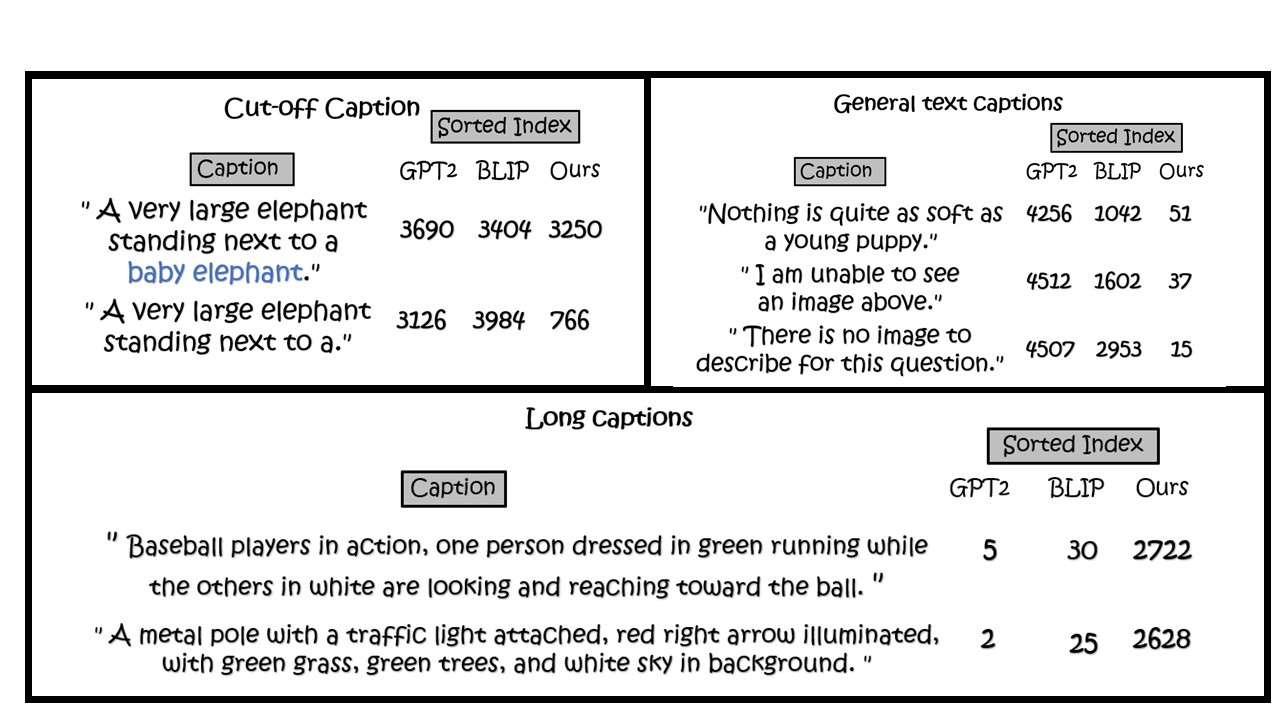}
    \caption{\textbf{Examples of differences between language models and our method.} The sorted index represents the position out of 5000 captions from MS-COCO, ranked from low to high likelihood values. The relative likelihood index is compared among GPT2 (LLM), BLIP (VLM), and our method. Top left: Our method shows a significant drop in relative likelihood compared to the language models when the sentence is cut-off. Top right: Captions that do not describe images receive the lowest relative likelihood from our method, the highest from GPT2, and intermediate scores from BLIP. Bottom: Long captions are assigned low relative likelihoods by language models, while our method assigns them average relative likelihood scores.}
    \label{fig: text diff exp}
\end{figure}

Figs.~\ref{fig: other text type}, \ref{fig: other text nouns} display histograms replicating the experiments from Figs.~\ref{fig: text diff}.a,b for additional models. The LLMs (OPT, NEO) exhibit behavior very similar to GPT-2. The VLMs (BLIP, GIT) also show behavior similar to GPT-2, but with some deviations trending toward our method's likelihood. This observation is reasonable, as these models, like CLIP, are trained (or fine-tuned) on caption data rather than general text data. In Fig.~\ref{fig: text diff exp}, we present additional examples of captions with varying relative likelihood scores. We sort the likelihood scores of 5,000 captions from MS-COCO in ascending order and examine the sorted index for different captions. The comparison includes our method, an LLM (GPT2), and a VLM (BLIP). Each set of examples demonstrates one of the three differences discussed in Sec.~\ref{sec: text model probabilities} (e.g. text type, grammatical errors and caption length).

The Area Under the Receiver Operating Characteristic Curve (AUC-ROC, AUC for simplicity) evaluates the ability of a model to distinguish between two classes. It measures the trade-off between the true positive rate (\( \text{TPR} \)) and the false positive rate (\( \text{FPR} \)) at various threshold levels. The AUC score in Tab.~\ref{table: text comp} is mathematically defined as:

\begin{equation}
    \text{AUC} = \int_{0}^{1} \text{TPR}(\text{FPR}^{-1}(x)) \, dx \quad , \quad \text{TPR} = \frac{\text{TP}}{\text{TP} + \text{FN}} \quad , \quad \text{FPR} = \frac{\text{FP}}{\text{FP} + \text{TN}}
    \label{eq: auc}
\end{equation}
where TP are true positives, FP are false positives, TN are true negatives and FN are false negatives. In the context of Tab.~\ref{table: text comp} the MS-COCO captions are defined as positives and the general text or captions without nouns are defined as negatives.

\section{Implementation and Theoretical Details}
\subsection{Implementation details}
\label{sec: imp det}
As explained in Sec.~\ref{sec: whiteneing transfrom} if $x$ is a random vector in $R^d$ with a non-singular covariance matrix $\Sigma$ (and with zero mean), then $W$ satisfying $W^TW = \Sigma^{-1}$ is called the
whitening matrix. One common approach to achieve whitening is through Principal Component Analysis (PCA), although other methods like Zero-Phase Component Analysis (ZCA) whitening and Singular Value Decomposition (SVD) whitening exist. PCA whitening transforms the data into a new coordinate system defined by the principal components of the covariance matrix. It rescales each component to have unit variance, effectively ``whitening'' the data. Steps of PCA whitening:

\begin{itemize}
    \item Compute the covariance matrix of the data.
    \item Perform eigenvalue decomposition to obtain eigenvalues (\( \Lambda \)) and eigenvectors (\( V \)).
    \item Transform the data:
    \begin{equation}
        \mathbf{X}_{\text{whitened}} = \Lambda^{-1/2} V^\top \mathbf{X}.
    \end{equation}
\end{itemize}

The main advantages of PCA whitening are that it ensures that the resulting features are uncorrelated and transforms data along principal axes, which often correspond to meaningful directions in the dataset. It can be efficient for dimensionality reduction, something we do not use in our work. The main limitation is the loss of original geometry (ZCA whitening for instance maintains the original geometry).

When the features in the original data are highly correlated, the matrix $W$ becomes unstable and may not be invertible. To address this issue, we remove one of the highly correlated features and replace it with random noise. In our experiments, this situation occurs only when whitening text embeddings and not with image embeddings. While this introduces some randomness into the process, it has minimal impact on the empirical results. Our full whitening code, together with scripts repeating our experiments is available  \href{https://github.com/rbetser/W_CLIP/tree/main}{here}.



\begin{algorithm}[ht]
\caption{Whitening Process}
\label{alg:pca_whitening}
\KwIn{Dataset \( \mathbf{X} \in \mathbb{R}^{N \times d} \), correlation threshold \( \tau \)}
\KwOut{Whitening matrix \( \mathbf{W} \)}

\BlankLine
\textbf{Step 1: Compute Correlation Matrix.} \\
\Indp
Calculate the correlation matrix:
\[
\mathbf{C}_{ij} = \frac{\text{Cov}(\mathbf{X}_i, \mathbf{X}_j)}{\sigma_i \sigma_j}.
\]
\Indm

\textbf{Step 2: Remove Highly Correlated Features.} \\
\Indp
Identify feature pairs \( (i, j) \) where \( |\mathbf{C}_{ij}| > \tau \).\\
For each pair, remove one feature (e.g., \( j \)) and replace it with random noise $\mathbf{r}$, Denote the updated dataset as \( \mathbf{X}' \):
\[
\mathbf{r} \sim \mathcal{N}(0, 0.1).
\]
\Indm

\textbf{Step 3: Compute Covariance Matrix.} \\
\Indp
Calculate the covariance matrix:
\[
\Sigma = \frac{1}{N} (\mathbf{X}'^\top \mathbf{X}').
\]
\Indm

\textbf{Step 4: Perform Eigenvalue Decomposition.} \\
\Indp
Decompose \( \Sigma \) into eigenvalues \( \Lambda \) and eigenvectors \( V \):
\[
\Sigma = V \Lambda V^\top.
\]
\Indm

\textbf{Step 5: Compute Whitening Matrix and Transform Data.} \\
\Indp
Calculate the whitening matrix:
\[
W = \Lambda^{-1/2} V^\top,
\]
\Indm

\Indp\Indp\Indp
where \( \Lambda^{-1/2} \) is a diagonal matrix with elements given by the inverse square root of the eigenvalues:
\[
\frac{1}{\sqrt{\lambda_i}}.
\]
\Indm
\end{algorithm}

\subsection{Isotropic random vectors}
\label{sec: isotropic}
An isotropic random vector is one where all components are identically distributed and statistically independent, with zero mean and unit variance, e.g it's covariance matrix $\mathbf{\Sigma}$ is the unit matrix. In other words, an isotropic vector is uniformly distributed across the space, exhibiting no directional bias. Such vectors often arise in high-dimensional statistical models and machine learning applications, where isotropy ensures that the data's statistical properties are invariant to rotation or translation \citep{mardia2000directional}. Isotropic distributions are particularly relevant in contexts such as embedding spaces, where uniformity and independence across features simplify analysis and facilitate probabilistic modeling \citep{vershynin2018high}.

\subsection{Normal distribution tests}
\label{sec: normal tests}
Normality tests assess whether a dataset follows a Normal distribution, a critical assumption in many statistical methods. As discussed above in Sec.~\ref{sec: Whitened CLIP Embeddings} the Anderson-Darling test evaluates how well the empirical cumulative distribution function (CDF) matches the expected CDF of a normal distribution, placing higher weight on the tails to detect deviations \citep{anderson1954test}. The D’Agostino-Pearson test combines skewness and shape characteristics measures to assess normality, offering sensitivity to both symmetric and asymmetric deviations \citep{d1973tests}. The Anderson-Darling test statistic is defined as:
\begin{equation}
    A^2 = -n - \frac{1}{n} \sum_{i=1}^{n} \left[ (2i - 1) \left( \ln F(y_i) + \ln \left(1 - F(y_{n+1-i}) \right) \right) \right]
    \label{eq: ad defin}
\end{equation}

Where \( n \) is the sample size, \( y_1 \leq y_2 \leq \dots \leq y_n \) are the ordered data samples and \( F(y) \) is the Cumulative Distribution Function (CDF) of the hypothesized distribution. The D’Agostino-Pearson test statistic combines skewness and kurtosis:

\begin{equation}
    K^2 = z_1^2 + z_2^2 \quad, \quad z_1 = \frac{g_1}{\sqrt{\frac{6}{n}}} \quad, \quad z_2 = \frac{g_2 - 3}{\sqrt{\frac{24}{n}}}
    \label{eq: dp defin}
\end{equation}

where \( K^2 \) is the D’Agostino-Pearson test statistic, \( z_1 \),  \( z_2 \) are the standardized skewness and kurtosis. \( g_1 \), \( g_2 \) are the sample skewness and kurtosis and \( n \) is the sample size. For details regarding skewness and kurtosis please refer to \citet{groeneveld1984measuring}.

\begin{figure}[H]
    \centering
    \includegraphics[width=0.95\textwidth]{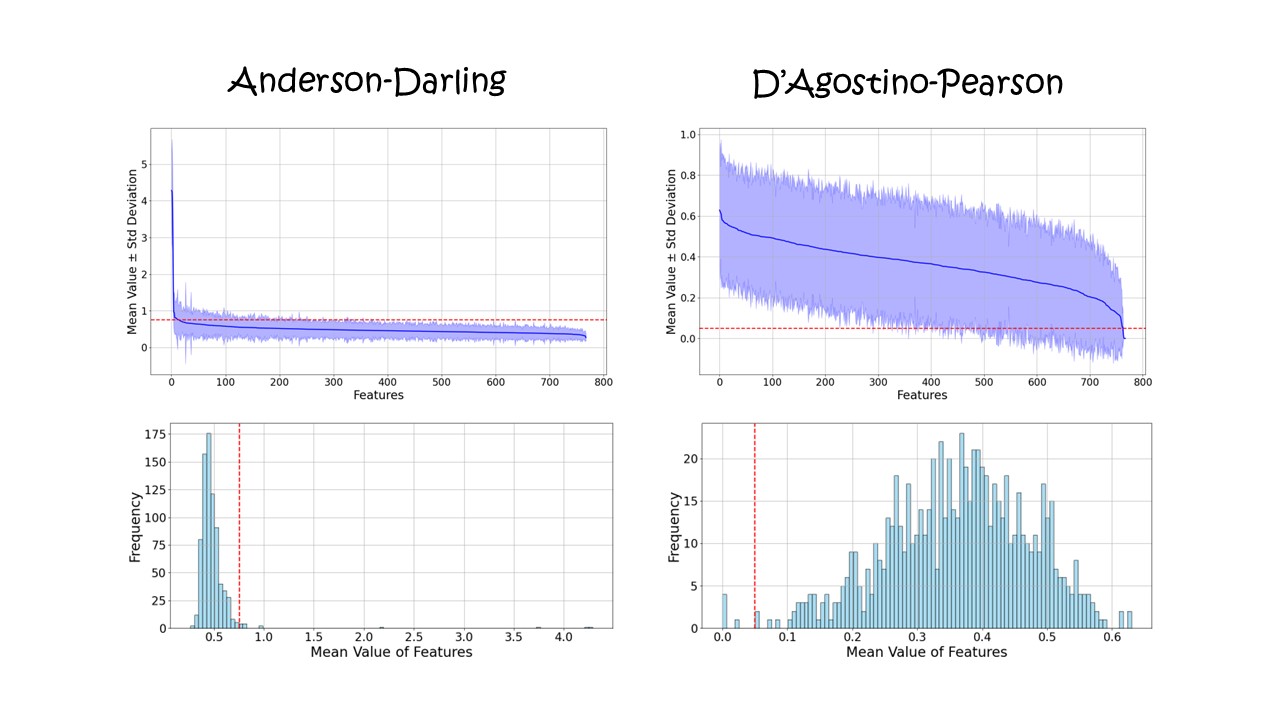}
    \caption{\textbf{Normal distribution tests on image embeddings.} Top row - mean value and standard deviation per feature, over all groups of embeddings. Bottom - histogram of mean values of each feature. In all plots the red line represents the test threshold. Left - Anderson-Darling test, threshold is 0.752, lower is better. Right - D’Agostino-Pearson test, threshold is 0.05, higher is better.}
    \label{fig: image gaus test}
\end{figure}

\begin{figure}[H]
    \centering
    \includegraphics[width=0.95\textwidth]{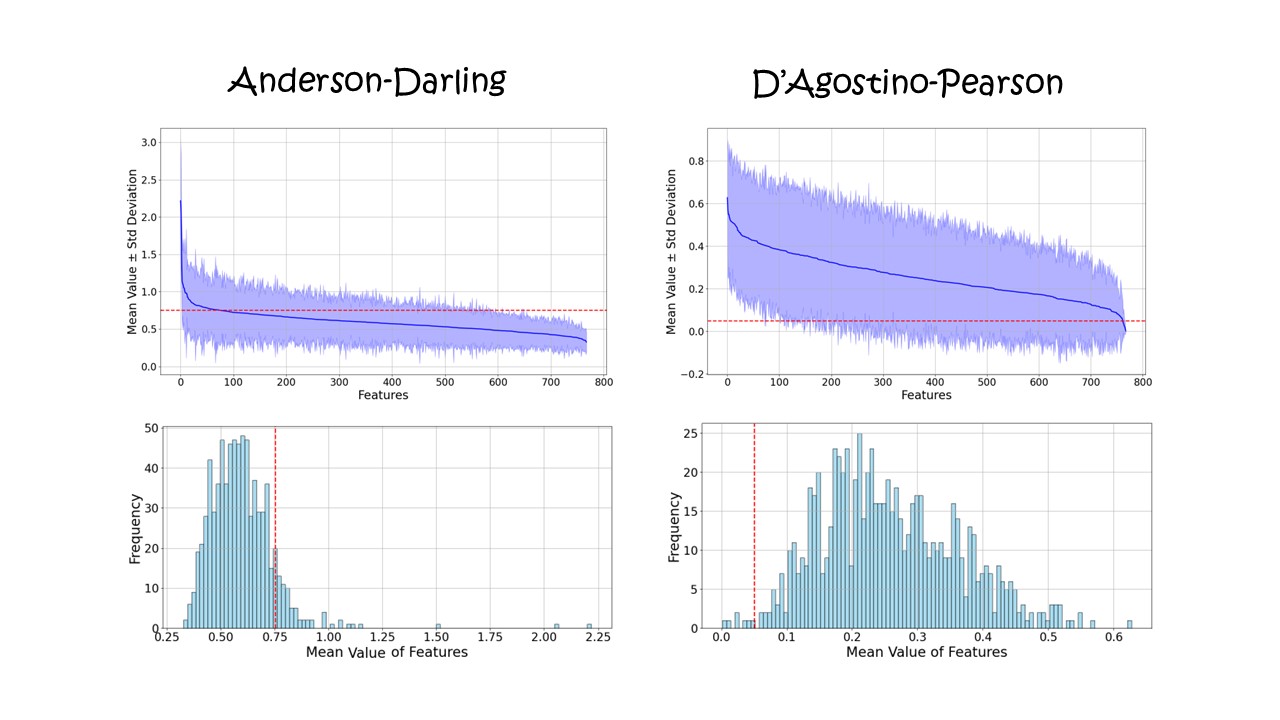}
    \caption{\textbf{Normal distribution tests on text embeddings.} Top row - mean value and standard deviation per feature, over all groups of embeddings. Bottom - histogram of mean values of each feature. In all plots the red line represents the test threshold. Left - Anderson-Darling test, threshold is 0.752, lower is better. Right - D’Agostino-Pearson test, threshold is 0.05, higher is better.}
    \label{fig: test gaus test}
\end{figure}

\begin{figure}[H]
    \centering
    \includegraphics[width=0.65\textwidth]{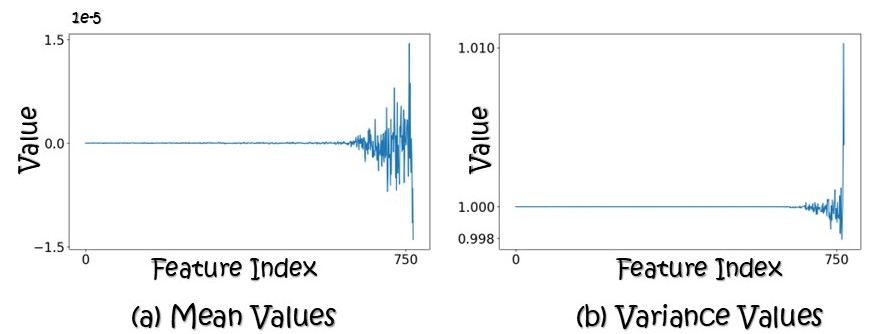}
    \caption{\textbf{Mean and variance of all whitened features.} We show the mean and variance of all the 768 features of the whitened embeddings. There are minor deviations from 0 (for mean) and 1 (for variance). For mean values the deviation is up to 0.0015\% and in the case of the variance the deviations are up to 1\%.}
    \label{fig: mean var}
\end{figure}

These tests are well-suited for high-dimensional data as they are robust to various types of distributional departures, making them effective for validating Normal approximations in the context of our proposed whitened embedding spaces. Bellow we present statistics of image (Fig.~\ref{fig: image gaus test}) and text (Fig.~\ref{fig: test gaus test}) embeddings, on both tests. Mean values (on all groups of data) with standard deviation and histograms of mean values are presented. In addition, in Fig.~\ref{fig: mean var} we present the mean and variance of all the whitened features, demonstrating minor deviation from the expected values (zero mean and unit variance).

\subsection{Thin shell theory}
\label{sec: thin shell}
The thin shell theory \citep{paouris2006concentration} is a concept in high-dimensional geometry. According to the thin shell theory most of the volume of a high dimensional convex space is concentered near the surface of the space. Specifically, for a convex space \( K \subseteq \mathbb{R}^d \), the majority of points in $K$ lie at an approximated distance $r$ from the origin. This can be formally described in terms of the concentration of measure, where the typical distance of a random point from the origin is concentrated around a specific radius:

\begin{equation}
    \mathbb{P}(\|x\| \in [r - \epsilon, r + \epsilon]) \approx \frac{C(d)}{\epsilon^d}
\label{eq: thin shell prob}
\end{equation}

where $x$ is a random sample from the space $K$, $r$ is the typical radius of the space, $\epsilon$ is a small deviation ($\epsilon \ll 1$) and $C$ is a constant that depends on the dimensions $d$. This result indicates that as the dimension \( d \) grows, the concentration near the thin shell becomes sharper. The thin shell phenomenon is closely related to the chi distribution described above (\eqnref{eq:log chi}). Specifically relating \eqnref{eq: chi mean std} to \eqnref{eq: thin shell prob} we get $r = \sqrt{d- \frac{1}{2}}$. Combining both phenomena, as $d$ increases, the space expands and concentration near the surface emerges because the majority of the space's mass resides near its boundary. Consequently, most points are located near the surface, even as the overall space grows.

\section{Ablation Study with Different Data and CLIP Model}
\label{sec: ablation}
We apply the whitening transform to embeddings of MS-COCO validation set using a second CLIP model - CLIP ViT-B/32, which encodes embeddings with 512 features (compared to embeddings with 768 features encoded by CLIP ViT-L/14). Results are in Tab.~\ref{table: clip b32}. Results are very similar to CLIP ViT-L-14, used in the paper, verifying that our method is general for different CLIP models. 

\begin{table}[H]
\centering
\caption{\textbf{Normal distribution tests using a different CLIP model} \textit{Avg. AD, DP} - the average Anderson-Darling, D’Agostino-Pearson p-value test scores (threshold is under 0.752 and above 0.05, respectively). MS-COCO validation set tested using CLIP ViT-B/32, which has 512 features in each embedding.}
\begin{tabular}{|c||cc|}
\hline
 & \textbf{Avg. AD} & \textbf{Avg. DP} \\ \hline \hline
\textbf{Image}        &   0.65 &  0.31  \\ 
\textbf{Text}         &   0.61 &  0.25 \\ \hline 
\end{tabular}
\label{table: clip b32}
\end{table}

\begin{figure}[H]
    \centering
    \includegraphics[width=0.75\textwidth]{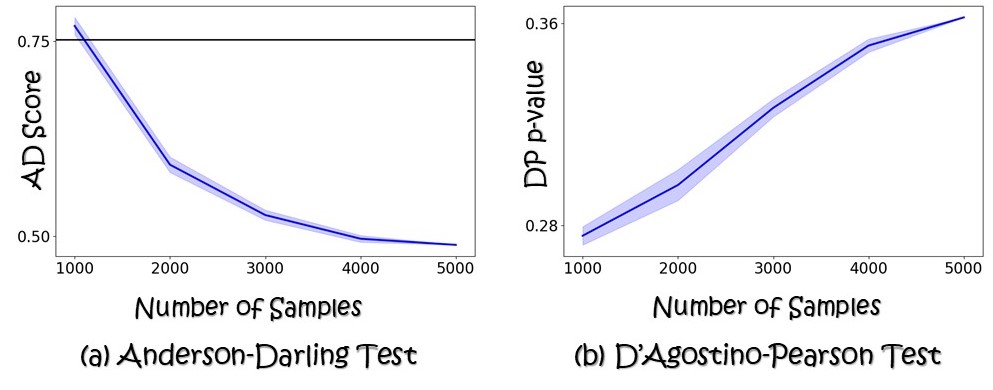}
    \caption{\textbf{Normal distribution tests with different data sizes}  Anderson-Darling average scores and standard deviation (a) and D’Agostino-Pearson p-value average scores and standard deviation (b). Threshold is under 0.752 (marked with a black line) and above 0.05, respectively. Computing the whitening matrix $W$ with different data sizes. For each size (1k, 2k, 3k, 4k) we randomly sampled 5 subsets of MS-COCO validation set and present the average score with standard deviation. The tests are performed on the full MS-COCO validation set (5k images).}
    \label{fig: ablation}
\end{figure}

We conduct an ablation study, examining the influence of the size of the data used for computing the whitening matrix $W$. For each size (1k, 2k, 3k, 4k) we randomly sampled 5 subsets of MS-COCO validation set. The average scores with standard deviation are plotted in Fig.~\ref{fig: ablation}. The tests are performed on the full MS-COCO validation set (5k images). For the D’Agostino-Pearson test, for all data sizes the tested embeddings comply with the normal distribution. For the Anderson-Darling test using 1k samples for computing $W$ results with tested embeddings that do not comply with the normal distribution. Using at least 2k results with tested embeddings that comply with the normal distribution. For both tests, using more data to compute $W$ results with improved results. As the whitening transform is completely data-driven it is expected that using additional data improves the results. However we note that the improvement between using 4k and 5k samples is small.

\section{Full circle SLERP Examples}
\label{sec: slerp exp}
\begin{figure}[ht]
    \centering
    \includegraphics[width=0.97\textwidth]{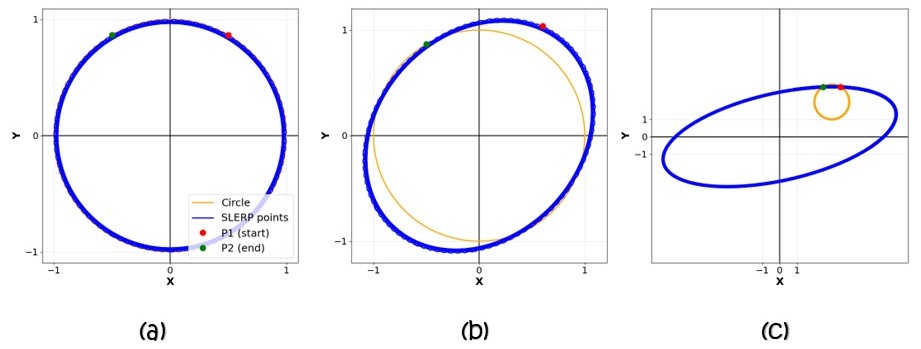}
    \caption{\textbf{2D full circle SLERP example.} The SLERP points are in blue and the circle perimeter is in orange. Examples of a simple 2D case of full circle SLERP. When both points are on the circle (a) the SLERP points follow the circle perimeter perfectly. If one of the points deviates from the circle (b) the SLERP points form an ellipse, but remain close to the circle perimeter. If the circle is skewed from the origin (c) the SLERP points form a large ellipse, that distances far from the circle perimeter.}
    \label{fig: slerp 2d}
\end{figure}

In Fig.~\ref{fig: slerp 2d} a simple 2D scenario of full circle SLERP is demonstrated. The main observation is that if the source and destination points are on a circle around the origin (allowing small deviations) the full circle SLERP points (blue) remain on (or near) the original circle (orange). However, if the circle is skewed from the origin the SLERP points deviate far from the original circle.

\begin{figure}[H]
    \centering
    \includegraphics[width=0.97\textwidth]{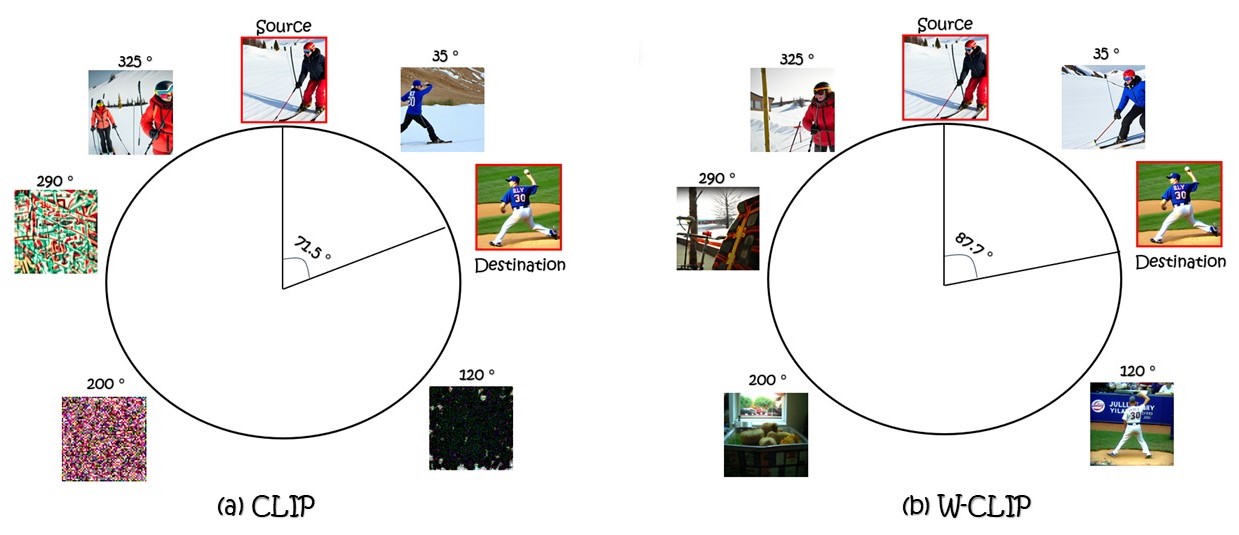}
    \caption{\textbf{Full circle SLERP example.} The full circle SLERP is performed in both the raw CLIP space (a) and in the W-CLIP space (b). The different angle between embeddings in both space is presented. In the raw CLIP space the full circle SLERP results with noise for most of the degrees not between the source and destination embeddings. In the W-CLIP space for all degrees real images are generated.}
    \label{fig: slerp far}
\end{figure}

We present an additional example of sets of images from a full circle SLERP, as discussed in Sec.~\ref{sec: image mani}, in Fig.~\ref{fig: slerp far}. As in Fig.~\ref{fig: slerp close}, also in this case it is clear that a full circle SLERP is not practical in the raw CLIP space, while resulting with real images throughout the full circle in the W-CLIP space. 

\begin{figure}[H]
    \centering
    \includegraphics[width=0.5\textwidth]{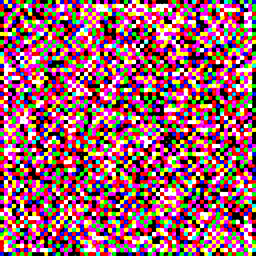}
    \caption{\textbf{Opposite image generated in the raw CLIP space.} The structured noise produced by CLIP exhibits 4×4 pixel blocks and a restricted color palette (black ('0' in all color channels), white ('1' in all color channels), red, green, blue, magenta ('1' in red and blue channels), cyan ('1' in green and blue channels), and yellow ('1' in red and green channels)), suggesting synthetic artifacts.}
    \label{fig: slerp opp}
\end{figure}

\newpage
\section{Image Generation Bias Examples}
\label{sec: clip loop exp}
\begin{figure}[ht]
    \centering
    \includegraphics[width=0.97\textwidth]{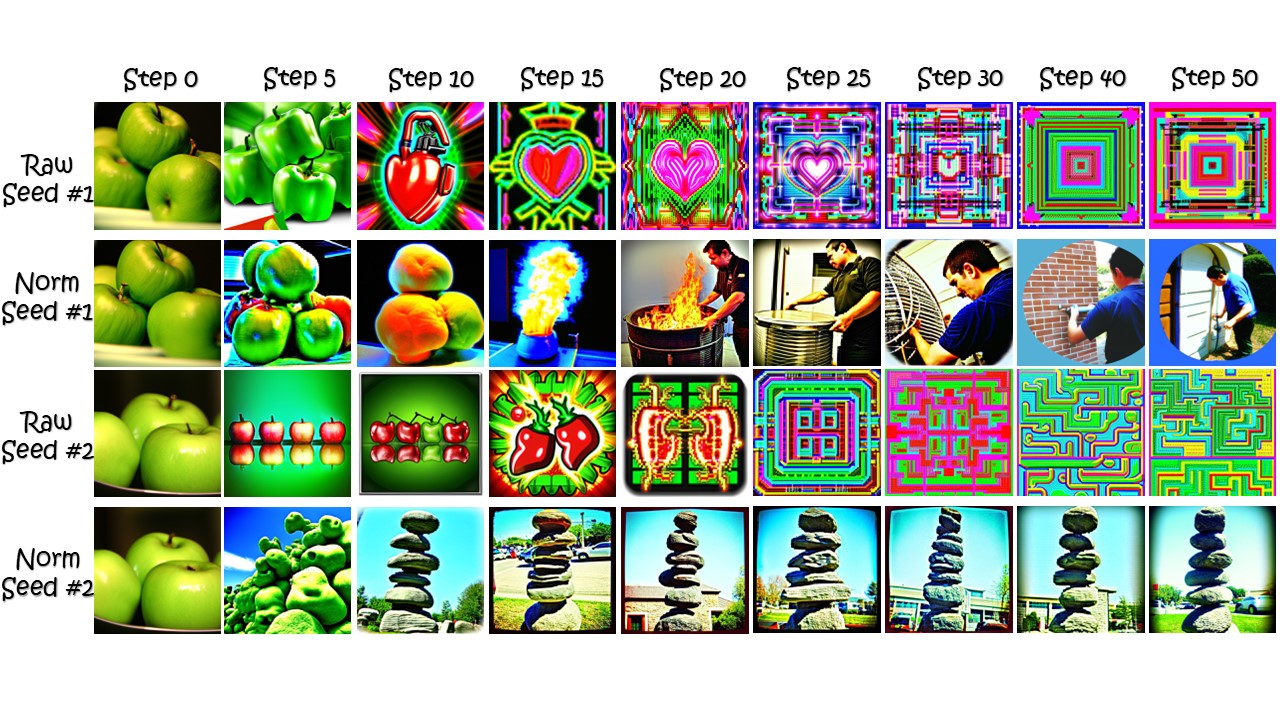}
    \caption{\textbf{Generation bias.} Iteratively using UnCLIP to generate images encoded by CLIP with two fixed seeds. The raw process gradually becomes noisy, whereas with normalization (to $\sqrt{d}$ at each encoding step), the content drifts but remains within a natural and reasonable image space.}
    \label{fig: clip loop apple}
\end{figure}

\begin{figure}[H]
    \centering
    \includegraphics[width=0.97\textwidth]{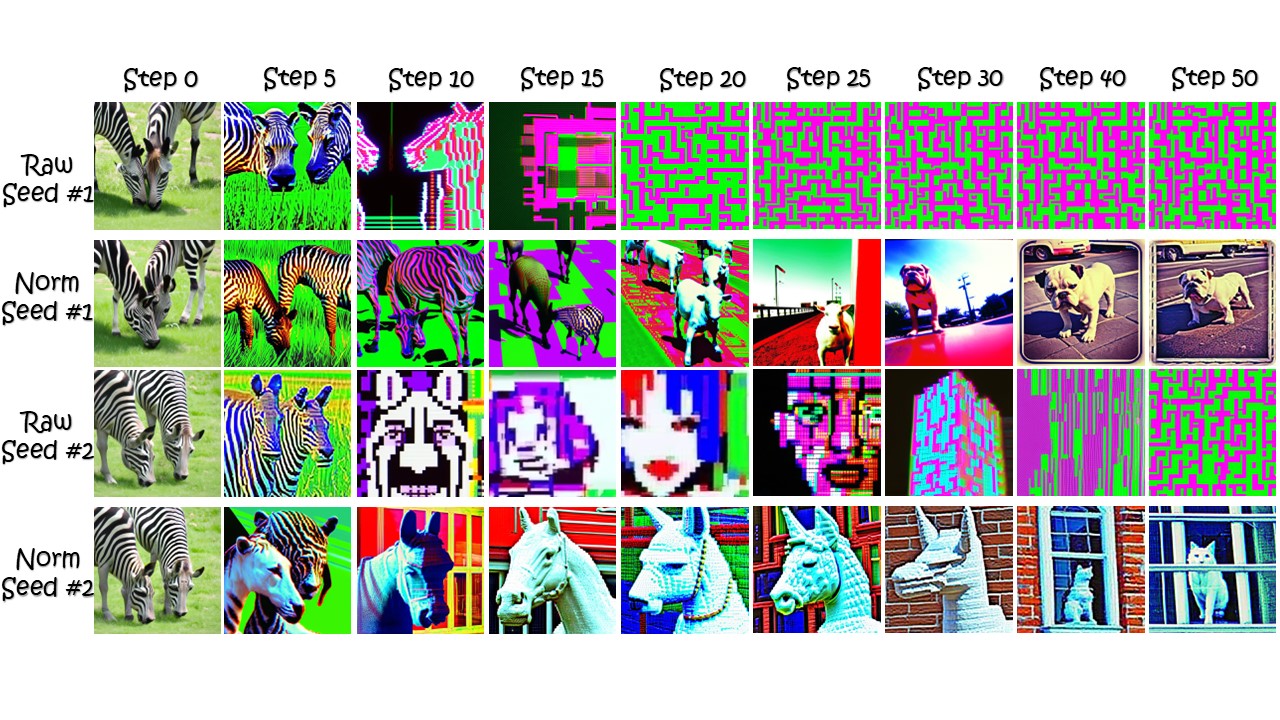}
    \caption{\textbf{Generation bias.} Iteratively using UnCLIP to generate images encoded by CLIP with two fixed seeds. The raw process gradually becomes noisy, whereas with normalization (to $\sqrt{d}$ at each encoding step), the content drifts but remains within a natural and reasonable image space.}
    \label{fig: clip loop zebra}
\end{figure}

In Figs.~\ref{fig: clip loop apple}, \ref{fig: clip loop zebra} additional examples of the bias in image generation models are presented. Different random seeds lead to different results due to the image generation model. In all cases the images become noise when no normalization is applied in the whitened space. Normalizing embeddings in the whitened space to have a norm of $\sqrt{d}$ in each iteration results with reasonable images, with varying content. 

\section{Text Complexity Examples}
\label{sec: detail text exp}
\begin{figure}[ht]
    \centering
    \includegraphics[width=0.95\textwidth]{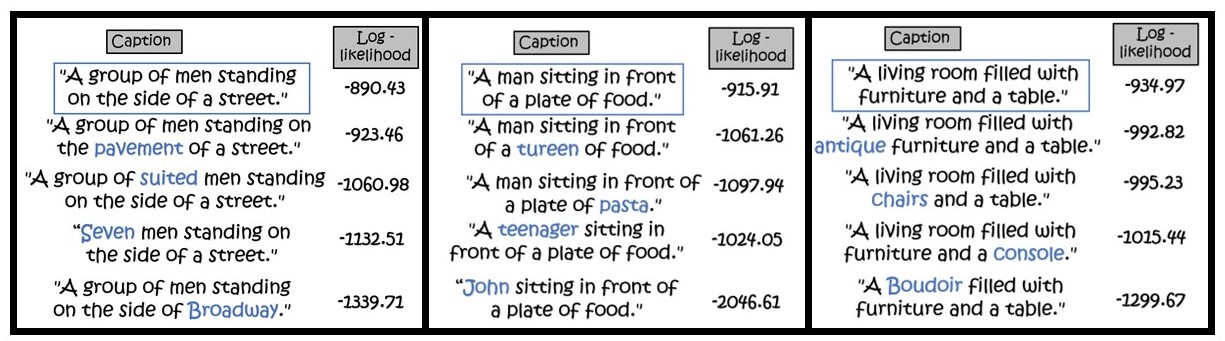}
    \caption{\textbf{Adding details to concepts.} The original caption from MS-COCO is framed in blue. Adding details decreases the likelihood.}
    \label{fig: add det}
\end{figure}

\begin{figure}[ht]
    \centering
    \includegraphics[width=0.95\textwidth]{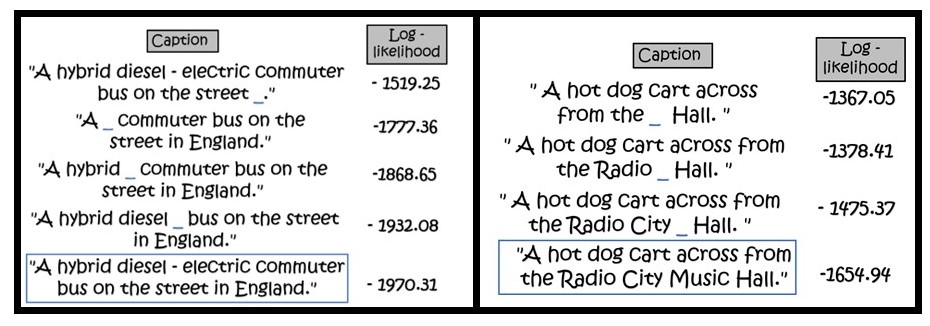}
    \caption{\textbf{Removing details from concepts.} The original caption from MS-COCO is framed with a blue frame. Removing different details increases the likelihood.}
    \label{fig: remove det}
\end{figure}

In Figs.~\ref{fig: add det}, \ref{fig: remove det} we repeat the experiment presented in Fig.~\ref{fig:text prob}, showing additional examples how adding and removing details from concepts (not the concepts themselves) decreases/increases the likelihood respectively.

\end{document}